\documentclass[acmsmall,screen,nonacm]{acmart}
\pdfoutput=1
\usepackage[utf8]{inputenc}
\usepackage[english]{babel}
\usepackage{booktabs}
\usepackage[capitalise,noabbrev,nameinlink]{cleveref}

\usepackage{subfig}
\usepackage{float}
\restylefloat{figure}
\usepackage{tabularx}

\usepackage{tikz}
\usetikzlibrary{backgrounds,calc,decorations,fit,shapes}
\usepackage{braket}
\usepackage[compat=0.6]{yquant}
\useyquantlanguage{groups}

\tikzset{%
ft/.style={%
  inner sep=2pt,
  font=\footnotesize
},
ft gate/.style={ft,shape=yquant-rectangle,rounded corners=2pt},
ft z/.style={ft,fill=fill-z,draw=draw-z},
ft x/.style={ft,fill=fill-x,draw=draw-x},
ft h/.style={ft,fill=fill-h,draw=draw-h},
ft bell/.style={ft,fill=fill-bell,draw=draw-bell},
}

\newlength{\xzskip}
\newcommand{\letteronx}[1]{%
\setbox0=\hbox{X}%
\setbox1=\hbox{#1}%
\setlength{\xzskip}{\wd0}%
\addtolength{\xzskip}{-\wd1}%
\phantom{X}\llap{#1\kern.5\xzskip}%
}
\newcommand{\zonx}{\letteronx{Z}}
\newcommand{\bonx}{\letteronx{B}}
\newcommand{\sonx}{\letteronx{S}}
\newcommand{\tonx}{\letteronx{T}}
\newcommand{\sdagonx}{\letteronx{S\smash{\kern-.5pt\lower1pt\hbox{$^\dagger$}}}}
\newcommand{\tdagonx}{\letteronx{T\smash{\kern-.5pt\lower1pt\hbox{$^\dagger$}}}}
\newcommand{\mvsrconx}{\letteronx{$\dashv$}}
\newcommand{\mvtgtonx}{\letteronx{$\vdash$}}

\yquantdefinebox{placeholder}[shape=yquant-circle,fill,radius=.5mm,anchor=center]{}

\yquantdefinebox{xgate}[ft gate,ft x]{X}
\yquantdefinebox{zgate}[ft gate,ft z]{\zonx}
\yquantdefinebox{tgate}[ft gate,ft z]{\tonx}
\yquantdefinebox{tdaggate}[ft gate,ft z]{\tdagonx}
\yquantdefinebox{hgate}[ft gate,ft h]{H}
\yquantdefinebox{sgate}[ft gate,ft h]{\sonx}
\yquantdefinebox{sdaggate}[ft gate,ft h]{\sdagonx}
\yquantdefinebox{ixgate}[ft gate,ft x]{$-i$X}
\yquantdefinebox{izgate}[ft gate,ft z]{$-i$Z}
\yquantdefinebox{movesrc}[shape=yquant-rectangle,rounded corners=2pt,ft bell]{\color{draw-bell}\mvsrconx}
\yquantdefinebox{movetgt}[shape=yquant-rectangle,rounded corners=2pt,ft bell]{\color{draw-bell}\mvtgtonx}

\yquantdefinebox{measx}[shape=yquant-ft-measure,ft x]{X}
\yquantdefinebox{measz}[shape=yquant-ft-measure,ft z]{\zonx}
\yquantdefinebox{measbell}[shape=yquant-ft-measure,ft bell]{\bonx}
\yquantdefinebox{prepbell}[shape=yquant-ft-prepare,ft bell]{\bonx}
\yquantdefinebox{prepx}[shape=yquant-ft-prepare,ft x]{X}
\yquantdefinebox{prepz}[shape=yquant-ft-prepare,ft z]{\zonx}
\yquantdefinebox{prept}[shape=yquant-ft-prepare,ft z]{\tonx}
\definecolor{fill-x}{RGB}{255,130,145}
\definecolor{draw-x}{RGB}{138,68,68}
\definecolor{fill-z}{RGB}{140,206,138}
\definecolor{draw-z}{RGB}{0,109,56}
\definecolor{fill-h}{RGB}{252,237,145}
\definecolor{draw-h}{RGB}{159,118,93}
\definecolor{fill-bell}{RGB}{228,146,55}
\definecolor{draw-bell}{RGB}{112,70,21}
\definecolor{griddark}{RGB}{142, 142, 142}
\definecolor{gridlight}{RGB}{235, 235, 235}

\newcommand{\ZCatCoord}[2]{%
\path[fill=fill-z,draw=draw-z,very thick] ($(#1)+(-45:2pt)$) arc (-45:225:2pt) -- ($(#2)+(135:2pt)$) arc (135:405:2pt) -- cycle;
}

\newcommand{\ZCat}[1]{%
\ZCatCoord{#1-0}{#1-p0}
}

\newcommand{\XCatCoord}[2]{%
\path[fill=fill-x,draw=draw-x,very thick] ($(#1)+(45:2pt)$) arc (45:315:2pt) -- ($(#2)+(225:2pt)$) arc (225:495:2pt) -- cycle;
}

\newcommand{\XCat}[1]{%
\path[fill=fill-x,draw=draw-x,very thick] ($(#1-0)+(-45:2pt)$) arc (-45:225:2pt) -- ($(#1-p0)+(135:2pt)$) arc (135:405:2pt) -- cycle;
}

\newcommand{\BellLine}[2]{%
\scoped[on background layer] \draw[draw-bell,line width=1mm] (#1) -- (#2);
}

\newcommand{\XLine}[2]{%
\scoped[on background layer]
\draw[draw-x,line width=1mm] (#1.center) -- (#2.center);
}

\newcommand{\ZLine}[2]{%
\scoped[on background layer]
\draw[draw-z,line width=1mm] (#1.center) -- (#2.center);
}

\newcommand{\XZLine}[2]{%
\begin{scope}[on background layer]
\draw[draw-x,line width=1mm] (#1.center) -- ($(#1.center)!.5!(#2.center)$);
\draw[draw-z,line width=1mm] (#2.center) -- ($(#2.center)!.5!(#1.center)$);
\end{scope}
}

\newcommand{\checkerboard}[2]{%
  \begin{scope}[on background layer]
  \foreach \row in {1,...,#1} {
    \foreach \col in {1,...,#2} {
      \pgfmathparse{int(mod(\col+\row,2))}
      \let\r\pgfmathresult
      \ifnum\r=0
      \fill[griddark] ($(\col,\row)-(1.5,1.5)$) rectangle ++(1,1);
      \else
      \fill[gridlight] ($(\col,\row)-(1.5,1.5)$) rectangle ++(1,1);
      \fi
    }
  }
  \draw[black] (-.5,-.5) rectangle ++(#2,#1);
  \end{scope}
}

\makeatletter
\DeclareRobustCommand\rvdots{%
\vbox{%
\baselineskip4\p@\lineskiplimit\z@%
\kern-\p@%
\hbox{.}\hbox{.}\hbox{.}%
}%
}
\makeatother

\makeatletter
\pgfdeclareshape{yquant-and}{%
   \saveddimen\xradius{%
      \pgf@x=.5\pgflinewidth%
   }%
   \saveddimen\yradius{%
      \pgfmathsetlength\pgf@x{\pgfkeysvalueof{/tikz/y radius}+.5*\yquant@config@register@sep}%
   }%
   \foreach \anc in {center, north, north east, east, south east, south, south west, west, north west} {%
      \inheritanchor[from=yquant-rectangle]{\anc}%
   }%
   \inheritanchorborder[from=yquant-rectangle]%
   \backgroundpath{%
      \pgf@xa=\dimexpr\yradius-\pgf@shorten@end@additional\relax%
      \pgf@ya=\dimexpr\yradius-\pgf@shorten@start@additional\relax%
      \pgf@shorten@end@additional=0pt %
      \pgf@shorten@start@additional=0pt %
      \pgfpathmoveto{\pgfqpoint{0pt}{\pgf@xa}}%
      \pgfpathlineto{\pgfqpoint{0pt}{0pt}}%
   }%
   \clippath{%
      \@tempdima=\dimexpr\yradius-\pgf@shorten@end@additional\relax%
      \@tempdimb=\dimexpr\yradius-\pgf@shorten@start@additional\relax%
      \pgfpathrectanglecorners%
         {\pgfqpoint{-.5\pgflinewidth}{\@tempdima}}%
         {\pgfqpoint{.5\pgflinewidth}{-\@tempdimb}}%
   }%
}
\makeatother

\makeatletter
\pgfdeclareshape{yquant-ft-prepare}{%
   \saveddimen\xradius{%
      \pgfmathsetlength\pgf@x{\pgfkeysvalueof{/tikz/x radius}}%
      \ifdim\wd\pgfnodeparttextbox=0pt %
         \pgf@xa=0pt %
      \else%
         \pgfmathsetlength\pgf@xa{\pgfkeysvalueof{/pgf/inner xsep}}%
      \fi%
      \ifdim\dimexpr.5\wd\pgfnodeparttextbox+\pgf@xa\relax>\pgf@x%
         \pgf@x=\dimexpr.5\wd\pgfnodeparttextbox+\pgf@xa\relax%
      \fi%
   }%
   \saveddimen\yradius{%
      \pgfmathsetlength\pgf@x{\pgfkeysvalueof{/tikz/y radius}}%
      \ifdim\dimexpr\ht\pgfnodeparttextbox+\dp\pgfnodeparttextbox\relax=0pt %
         \pgf@xa=0pt %
      \else%
         \pgfmathsetlength\pgf@xa{\pgfkeysvalueof{/pgf/inner ysep}}%
      \fi%
      \@tempdima=\dimexpr.5\ht\pgfnodeparttextbox+.5\dp\pgfnodeparttextbox+\pgf@xa\relax%
      \ifdim\@tempdima>\pgf@x%
         \pgf@x=\@tempdima%
      \fi%
   }%
   \inheritsavedanchors[from=yquant-rectangle]%
   \foreach \anc in {text, center, north, north east, east, south east, south, south west, west, north west} {%
      \inheritanchor[from=yquant-rectangle]{\anc}%
   }%
   \inheritanchorborder[from=yquant-rectangle]%
   \backgroundpath{%
      \pgfpathmoveto{\pgfpoint{0pt}{-\yradius}}%
        \pgfpathlineto{\pgfpoint{\xradius}{-\yradius}}%
        \pgfpathlineto{\pgfpoint{\xradius}{\yradius}}%
        \pgfpathlineto{\pgfpoint{0pt}{\yradius}}%
        \pgfpathcurveto{\pgfpoint{.8*-\dimexpr\xradius}{\yradius}}{\pgfpoint{-\xradius}{.8\dimexpr\yradius}}{\pgfpoint{-\xradius}{0}}
        \pgfpathcurveto{\pgfpoint{-\xradius}{-.8\dimexpr\yradius}}{\pgfpoint{.8*-\dimexpr\xradius}{-\yradius}}{\pgfpoint{0}{-\dimexpr\yradius}}%
   }%
   \inheritclippath[from=yquant-rectangle]%
}
\pgfdeclareshape{yquant-ft-measure}{%
   \saveddimen\xradius{%
      \pgfmathsetlength\pgf@x{\pgfkeysvalueof{/tikz/x radius}}%
      \ifdim\wd\pgfnodeparttextbox=0pt %
         \pgf@xa=0pt %
      \else%
         \pgfmathsetlength\pgf@xa{\pgfkeysvalueof{/pgf/inner xsep}}%
      \fi%
      \ifdim\dimexpr.5\wd\pgfnodeparttextbox+\pgf@xa\relax>\pgf@x%
         \pgf@x=\dimexpr.5\wd\pgfnodeparttextbox+\pgf@xa\relax%
      \fi%
   }%
   \saveddimen\yradius{%
      \pgfmathsetlength\pgf@x{\pgfkeysvalueof{/tikz/y radius}}%
      \ifdim\dimexpr\ht\pgfnodeparttextbox+\dp\pgfnodeparttextbox\relax=0pt %
         \pgf@xa=0pt %
      \else%
         \pgfmathsetlength\pgf@xa{\pgfkeysvalueof{/pgf/inner ysep}}%
      \fi%
      \@tempdima=\dimexpr.5\ht\pgfnodeparttextbox+.5\dp\pgfnodeparttextbox+\pgf@xa\relax%
      \ifdim\@tempdima>\pgf@x%
         \pgf@x=\@tempdima%
      \fi%
   }%
   \inheritsavedanchors[from=yquant-rectangle]%
   \foreach \anc in {text, center, north, north east, east, south east, south, south west, west, north west} {%
      \inheritanchor[from=yquant-rectangle]{\anc}%
   }%
   \inheritanchorborder[from=yquant-rectangle]%
   \backgroundpath{%
      \pgfpathmoveto{\pgfpoint{0pt}{-\yradius}}%
        \pgfpathlineto{\pgfpoint{-\xradius}{-\yradius}}%
        \pgfpathlineto{\pgfpoint{-\xradius}{\yradius}}%
        \pgfpathlineto{\pgfpoint{0pt}{\yradius}}%
        \pgfpathcurveto{\pgfpoint{.8*\dimexpr\xradius}{\yradius}}{\pgfpoint{\xradius}{.8\dimexpr\yradius}}{\pgfpoint{\xradius}{0}}
        \pgfpathcurveto{\pgfpoint{\xradius}{-.8\dimexpr\yradius}}{\pgfpoint{.8*\dimexpr\xradius}{-\yradius}}{\pgfpoint{0}{-\dimexpr\yradius}}%
   }%
   \inheritclippath[from=yquant-rectangle]%
}
\makeatother

\newcommand{\todo}[1]{\textcolor{red}{[TODO:] #1}}
\renewcommand{\todo}[1]{}

\acmDOI{00.000/000_0}
\acmISBN{000-0000-00-000/00/00}
\acmConference[ABC '00]{Marine Science Center End Of Summer Conference}{August 0000}{Bamfield, BC, Canada}
\acmYear{0000}
\copyrightyear{0000}
\acmPrice{0.00}
\setcopyright{none}

\usepackage{natbib}
\usepackage{graphicx}

\usetikzlibrary{external}
\tikzexternaldisable

\newcommand{\inputtikz}[1]{%
\includegraphics{tikz/#1}}

\begin{abstract}
	We describe a space-time optimized circuit for the table lookup subroutine from lattice-surgery surface code primitives respecting 2D grid connectivity. Table lookup circuits are ubiquitous in quantum computing, allowing the presented circuit to be used for applications ranging from cryptography to quantum chemistry. Surface code is the leading approach to scalable fault-tolerant quantum computing pursued by industry and academia.  We abstract away surface code implementation details by using a minimal set of operations supported by the surface code via lattice-surgery. Our exposition is accessible to a reader not familiar with surface codes and fault-tolerant quantum computing.
\end{abstract}

\begin{document}

\title{Space-time optimized table lookup}
\author{Thomas H\"aner}
\authornote{This work was completed prior to T.H. joining AWS}
\affiliation{%
\institution{Microsoft}
\country{Switzerland}
}
\author{Vadym Kliuchnikov}
\affiliation{%
\institution{Microsoft}
\country{Canada}
}
\author{Martin Roetteler}
\affiliation{%
\institution{Microsoft}
\country{United States}
}
\author{Mathias Soeken}
\affiliation{%
\institution{Microsoft}
\country{Switzerland}
}

\maketitle


\section{Introduction}
\label{sec:intro}

For certain computational tasks, quantum computers are known to asymptotically outperform traditional computers running state-of-the-art algorithms. This includes factoring~\cite{Shor97} and simulating systems that must be treated quantum mechanically to achieve the desired accuracy~\cite{BLH+20}. To assess whether an asymptotic speedup translates to a computational advantage \emph{in practice}, a quantum algorithm must be broken down into elementary operations for which runtime estimates can be derived from the chosen error correction protocol.
Promising applications can then be optimized to reduce the resource requirements (qubits and time) further. In turn, this reduces the time until the given quantum algorithm can be used to achieve a quantum advantage at application scale.

Over the recent years, promising quantum algorithms for applications such as factoring and simulating quantum chemistry have been optimized significantly~\cite{GE19,BLH+20,LBG+21,HJN+20}. To this end, researchers have explored various tradeoffs (e.g., space for time) in order to reduce the many overheads that must be introduced to achieve a fault-tolerant implementation.

The table lookup is one subroutine that can be used to reduce the resource requirements for state preparation~\cite{LKS18} and arithmetic~\cite{GE19}. As such, it is used in state-of-the-art implementations of both Shor's algorithm for factoring and computing discrete logarithms~\cite{GE19,HJN+20} and algorithms for solving chemistry problems~\cite{BLH+20,LBG+21}. Given this broad range of applications, it makes sense to manually optimize this subroutine, including a detailed layout for execution based on lattice-surgery surface code.
We explore various implementation tradeoffs and assess their impact on the total physical space/time volume under different assumptions regarding the fault-tolerance protocol and qubit quality.

The paper is structured as follows.  The next section introduces the table logic operation on a logical level.  \cref{sec:mapping} describes how to map the state-of-the-art table lookup construction for table lookup to surface code and derives general cost estimates in terms of qubit count logical compute cycle count.  \cref{sec:tradeoff} proposes a new construction that allows to reduce the compute cycle count by using more qubits.  \cref{sec:gadgets} describes several universally useful gadgets that are used throughout the paper.

\section{Table lookup}
Here, we briefly introduce the basic quantum operations (gates) that we use to
implement table lookup. For a thorough introduction to quantum computing, we
refer the reader to the textbook by Nielsen and Chuang~\cite{NC00}.

The controlled NOT or \texttt{CNOT} gate, with the first qubit acting as the control qubit, is defined as 
\begin{equation}
    |c\rangle|t\rangle \mapsto |c\rangle|t\oplus c\rangle,
\end{equation}
and the multi-target \texttt{CNOT} gate is defined as
\begin{equation}
    |c\rangle|t_1\rangle\ldots|t_m\rangle \mapsto |c\rangle|t_1\oplus c\rangle \ldots|t_m\oplus c\rangle,
\end{equation}
where we use `$\oplus$' to denote both Boolean and bitwise XOR.

Given $K$ bit strings $d_0, \dots, d_{K-1}$, each of length $m$, a table lookup with
$k \ge \lceil\log_2 K\rceil$ input qubits $|x\rangle = |x_1 \dots x_k\rangle$
and $m$ output qubits $|y\rangle = |y_1\dots y_m\rangle$ maps
\begin{equation}
    |x\rangle |y\rangle \mapsto |x\rangle |y \oplus f(x)\rangle,
\end{equation}
where $f(x_1, \dots, x_k) = d_x$, if $x = (x_1 \dots x_k)_2 < K$, and an arbitrarily chosen output if $x\geq K$~\cite{BGB+18} (for optimization purposes).  In the input
assignment, $x_1$ is the most significant bit. We assume that the circuit will
never be run on undefined inputs, or that undefined outputs do not negatively
affect the overall algorithm.  For $K=1$ and $k=0$ input qubits, the function
$f$ turns into the constant bit string $d_0$, and the definition becomes
$|y\rangle \mapsto |y\oplus d_0\rangle$.  We say that $d_0$ is the \emph{data
being looked up} by the table lookup circuit.  The controlled table lookup
function is defined similarly with an additional control qubit $|c\rangle$.
\begin{equation}
    |c\rangle|x\rangle|y\rangle
    \mapsto
    |c\rangle|x\rangle|c\mathbin{?}y\oplus f(x)\mathbin{:}y\rangle,
\end{equation}
where the `$\cdot \mathbin{?}\cdot\mathbin{ : }\cdot$' denotes the
\emph{if-then-else} operation.

The unary-iterate table lookup construction~\cite{BGB+18} requires $O(K)$ $T$
gates, in particular, the number does not depend on the number of output
bits~\cite{BGB+18}.  We will first illustrate the construction for a single
input, i.e., $k=1$, and then derive a recursive construction for the general
case.

\begin{figure}[t]
    \centering
    \footnotesize
    \inputtikz{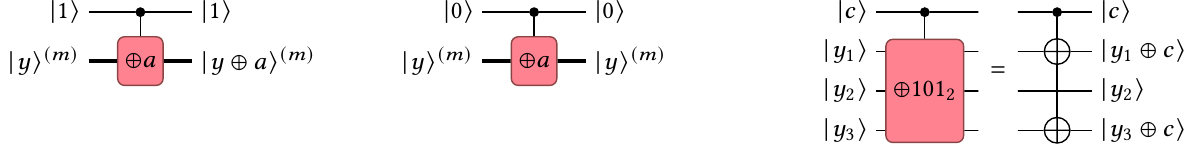}
    \caption[Bitwise XOR gates]{\texttt{CXOR} gates.  A bold line represents a
    multi-qubit register, in which the $^{(m)}$ suffix to a qubit register
    denotes the number of bits in the register.}
    \label{fig:cxor}
\end{figure}

\paragraph*{One-input table lookup.}
For any bit string $a$ of length $m$, the bitwise \texttt{XOR} operation maps
\begin{equation}
    |y\rangle \mapsto |y \oplus a\rangle,
\end{equation}
where $|y\rangle$ is an $m$-qubit computational basis state.  One can think of this operation as a table lookup with zero inputs.  Similarly, its controlled version  can be seen as a controlled table lookup with zero inputs
\begin{equation}
|c\rangle|y\rangle \mapsto |c\rangle|c \mathbin{?} y \oplus a \mathbin{:} y\rangle.
\end{equation}
This controlled bitwise \texttt{XOR} operation, called \texttt{CXOR}, can be
implemented using multi-target \texttt{CNOT} gates, see \cref{fig:cxor}.

A one-input table lookup for two bit strings $d_0$ and $d_1$ than can be
implemented as
\begin{equation}
\inputtikz{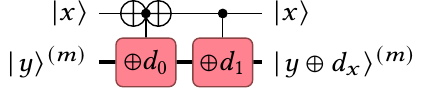}
\end{equation}
for a function $f : \{0, 1\} \to \{0, 1\}^m$.  The bold line and $^{(m)}$-suffix represent an $m$-qubit register and the output is any bit string of length $m$.  Note that a one-input table lookup circuit requires only Clifford operations and a zero-input table lookup requires only Pauli gates.

\begin{figure}[t]
    \centering
    \inputtikz{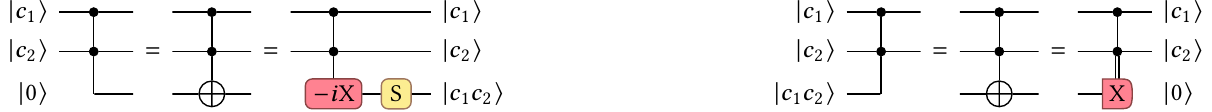}    
    \caption[\texttt{AND} and \texttt{AND}$^{\dagger}$ gates]{ \texttt{AND} and \texttt{AND}$^{\dagger}$ gates. Each of the variables $c_1$, $c_2$ is represented by a single qubit. }
    \label{fig:and-gate}
\end{figure}

To make an efficient circuit for a controlled one-input  table lookup, we make
use of \texttt{AND} and \texttt{AND}$^{\dagger}$ gates as defined in
\cref{fig:and-gate}. We use the shorthand $c_1c_2$ to denote the AND of the bits
$c_1$ and $c_2$. Note that the \texttt{AND} gate requires four $|T\rangle$ magic
states or one $|\mathrm{CCZ}\rangle$ magic state, whereas
\texttt{AND}$^{\dagger}$ requires only Clifford operations.  Note how the
\texttt{AND} gate in~\cref{fig:and-gate} is expressed in terms of
a~\texttt{CCiX} gate and an $S$ gate on the target qubits.

Using one \texttt{AND} and one \texttt{AND}$^{\dagger}$ gate, the controlled one-input  table lookup can be implemented as
\begin{equation}
  \label{eq:controlled-one-input-table-lookup}
  \inputtikz{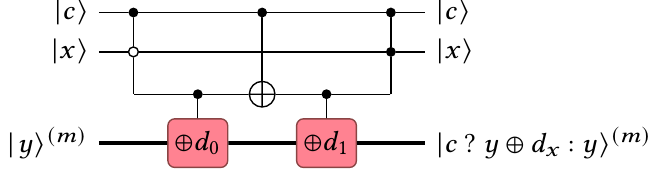}
\end{equation}
where $f: \{0, 1\} \to \{0, 1\}^m$. The first \texttt{CXOR} gate is executed
when $|c\rangle$ is one and $|x\rangle$ is zero. The second \texttt{CXOR} gate
is executed when both $|c\rangle$ and $|x\rangle$ are one.

\paragraph*{Recursive construction}
It is more convenient to first construct a \textit{controlled} table lookup recursively to then use this construction to arrive at an implementation of a non-controlled table lookup.

\begin{figure}[t]
    \centering
    \footnotesize
    \subfloat[\label{fig:controlled-table-lookup}]{\inputtikz{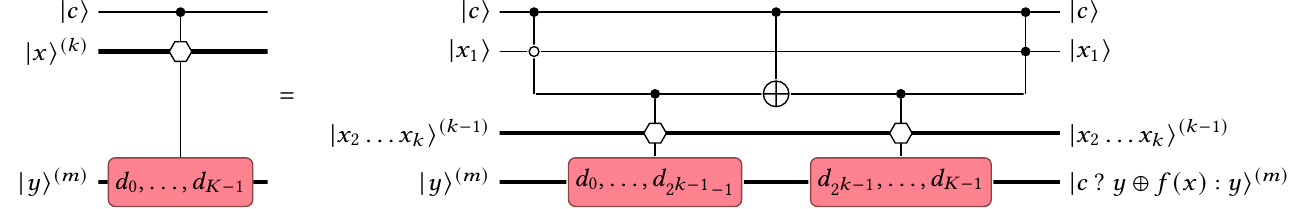}}

    \subfloat[\label{fig:unary-iterate-table-lookup}]{\inputtikz{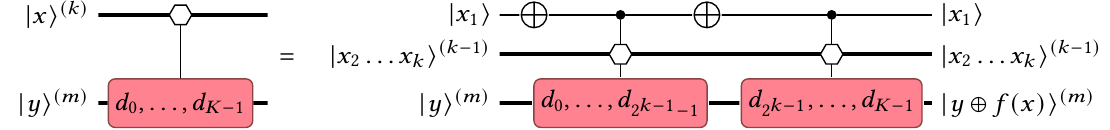}}
    \caption{(a) controlled and (b) uncontrolled $k$-input and $m$-output table lookup circuit; we use a hexagon to denote the input register for the table lookup.}
    \label{fig:controlled-table-lookup-tmp}
\end{figure}

The controlled table lookup construction for $K$ bit strings $d_0, \dots,
d_{K-1}$ and $k = \lceil\log_2 K\rceil$ inputs is depicted in
\cref{fig:controlled-table-lookup}.  Based on the controlled one-input table
lookup in Eq.~\eqref{eq:controlled-one-input-table-lookup}, we construct a
controlled $k$-input table lookup by splitting the bit strings into two sets,
the first one containing the first $2^{k-1}$ bit strings, and the second one
containing the rest.  Note that the first set is addressed when the
most-significant bit $x_1$ is 0, and the second set is addressed when $x_1$ is
1.  This corresponds to the two co-factors $f(0, x_2, \dots, x_k)$ and $f(1,
x_2, \dots, x_k)$ of $f$.  The input register in the second controlled table
lookup in~\cref{fig:controlled-table-lookup} on the right-hand side might
require fewer than $k - 1$ inputs.

\cref{fig:unary-iterate-table-lookup} shows how to combine two controlled
$(k-1)$-input table lookup circuits to arrive at a $k$-input table lookup circuit. We
first apply the negatively controlled version to look up all bit strings at
addresses starting with zero and then we apply the controlled version to look up
all bit strings at addresses starting with a one. Note that we could also switch
the order of negatively and positively controlled table lookup circuits. We will
show later that they can also be run in parallel to increase the magic state consumption rate and the utilization of
output qubits.

\paragraph*{Magic state requirements}
Let $N^{\text{TL}}_{\text{AND}}(k)$ be the number of \texttt{AND} gates used 
by table lookup with $k$ inputs and  $N^{\text{CTL}}_{\text{AND}}(k)$
be the number of \texttt{AND} gates used 
by controlled table lookup with $k$ inputs.

Then we have the following: 
\begin{align}
   N^{\text{TL}}_{\text{AND}}(k) & = 2N^{\text{CTL}}_{\text{AND}}(k-1) \\
   N^{\text{CTL}}_{\text{AND}}(0) & = 0 \\
   N^{\text{CTL}}_{\text{AND}}(1) & = 1 \\
   N^{\text{CTL}}_{\text{AND}}(k) & = 1 + 2 N^{\text{CTL}}_{\text{AND}}(k-1) = 2^{k} - 1, \text{ for } k \ge 0 \\
   N^{\text{TL}}_{\text{AND}}(k) & = 2 \cdot (2^{k-1} - 1) = 2^{k} - 2, \text{ for } k \ge 1 \\
   N^{\text{TL}}_{\text{AND}}(0) & = 0 
\end{align}
Therefore the number of T states required by unary-iterate table lookup 
is
\begin{align}
   N^{\text{TL}}_{\text{T}}(k) & = 4(2^{k} - 2) = 2^{k+2}-8, \text{ for } k \ge 1 \\
   N^{\text{TL}}_{\text{T}}(0) & = 0
\end{align}
\section{Mapping to lattice-surgery surface code}
\label{sec:mapping}

\begin{table}
    \centering
    \caption{Operations supported by surface codes with lattice surgery and restrictions on them. Gray circles indicate that operation uses neighbouring qubits; the qubits start and end in a ``blank'' state.}
    \label{fig:surface-code-operations}
\tikzexternaldisable
\begin{tabular}{llll}
\toprule
\multicolumn{2}{l}{Symbol} & Name & Remarks \\
\midrule
\small Circuit & \small 2D & & \\[4pt]
\begin{tikzpicture}[baseline=(base)]
\begin{yquant}
qubit {} q;
[font=\protect\normalsize] measz q;
discard q;
\end{yquant}
\coordinate (base) at (current bounding box.center);
\end{tikzpicture}
&
\scalebox{.8}{\input{circuits/mini-measz}}
& Measure Z & Leaves qubit in a ``blank'' state \\[10pt]
\begin{tikzpicture}[baseline=(base)]
\begin{yquant}
qubit {} q;
[font=\protect\normalsize]
measx q;
discard q;
\end{yquant}
\coordinate (base) at (current bounding box.center);
\end{tikzpicture}
&
\scalebox{.8}{\input{circuits/mini-measx}}
& Measure X & Leaves qubit in a ``blank'' state \\[10pt]
\begin{tikzpicture}[baseline=(base)]
\begin{yquant}
qubit {} q;
setstyle {draw=none} -;
[font=\protect\normalsize]
prepz q;
setstyle {draw} -;
\end{yquant}
\coordinate (base) at (current bounding box.center);
\end{tikzpicture}
&
\scalebox{.8}{\input{circuits/mini-prepz}}
& Prepare $|0\rangle$ & Expects qubit in a ``blank'' state \\[10pt]
\begin{tikzpicture}[baseline=(base)]
\begin{yquant}
qubit {} q;
setstyle {draw=none} -;
[font=\protect\normalsize]
prepx q;
setstyle {draw} -;
\end{yquant}
\coordinate (base) at (current bounding box.center);
\end{tikzpicture}
&
\scalebox{.8}{\input{circuits/mini-prepx}}
& Prepare $|+\rangle$ & Expects qubit in a ``blank'' state \\[10pt]
\begin{tikzpicture}[baseline=(base)]
\begin{yquant}
qubit {} q[2];
[name=z1, font=\protect\normalsize]
measz q[0];
[name=z2, font=\protect\normalsize]
measz q[1];
\end{yquant}
\ZLine{z1}{z2}
\coordinate (base) at (current bounding box.center);
\end{tikzpicture}
&
\scalebox{.8}{
\begin{tikzpicture}[x=.5cm,y=-.5cm,baseline=(base)]
    \checkerboard{2}{2}
    
    \begin{scope}[every node/.style={inner sep=1.5pt,font=\normalsize}]
    \node[shape=yquant-ft-measure,ft z] (zz-a) at (0,0) {\zonx};
    \node[shape=yquant-ft-measure,ft z] (zz-b) at (0,1) {\zonx};
    \end{scope}
    \ZLine{zz-a}{zz-b}
    
    \coordinate (base) at (current bounding box.center);
\end{tikzpicture}}
& Measure ZZ & Vertical only \\[10pt]
\begin{tikzpicture}[baseline=(base)]
\begin{yquant}
qubit {} q[2];
[name=x1, font=\protect\normalsize]
measx q[0];
[name=x2, font=\protect\normalsize]
measx q[1];
\end{yquant}
\XLine{x1}{x2}
\coordinate (base) at (current bounding box.center);
\end{tikzpicture}
&
\scalebox{.8}{
\begin{tikzpicture}[x=.5cm,y=-.5cm,baseline=(base)]
    \checkerboard{2}{2}
    
    \begin{scope}[every node/.style={inner sep=1.5pt,font=\normalsize}]
    \node[shape=yquant-ft-measure,ft x] (xx-a) at (0,0) {X};
    \node[shape=yquant-ft-measure,ft x] (xx-b) at (1,0) {X};
    \end{scope}
    \XLine{xx-a}{xx-b}
    
    \coordinate (base) at (current bounding box.center);
\end{tikzpicture}}
& Measure XX & Horizontal only \\[10pt]
\begin{tikzpicture}[baseline=(base)]
\begin{yquant}
qubit {} q[2];
[name=x1, font=\protect\normalsize]
measx q[0];
[name=z2, font=\protect\normalsize]
measz q[1];
\end{yquant}
\XZLine{x1}{z2}
\coordinate (base) at (current bounding box.center);
\end{tikzpicture}
&
\scalebox{.8}{
\begin{tikzpicture}[x=.5cm,y=-.5cm,baseline=(base)]
    \checkerboard{2}{2}
    
    \begin{scope}[every node/.style={inner sep=1.5pt,font=\normalsize}]
    \node[shape=yquant-ft-measure,ft x] (xx) at (0,0) {X};
    \node[shape=yquant-ft-measure,ft z] (zz) at (1,1) {\zonx};
    \end{scope}
    
    \path[draw,fill=white!90!gray] (0,1) circle (.3);
    \path[draw,fill=white!90!gray] (1,0) circle (.3);
    
    \XZLine{xx}{zz}
    
    \coordinate (base) at (current bounding box.center);
\end{tikzpicture}}
& Measure XZ & Blocks neighbors; expects and leaves them in ``blank'' state \\
\bottomrule
\end{tabular}
\tikzexternalenable
\end{table}

The goal of this section is to describe an explicit mapping of table lookup to
lattice surgery operations on logical qubits in a 2D grid.  We do this by using
the basic operations given in~\cref{fig:surface-code-operations} and assuming
the availability of $|S\rangle$- and $|T\rangle$-states.  We also use these
operations to construct more high-level gadgets described
in~\cref{sec:gadgets}.

We start by applying several circuit
transformations and optimizations to the table lookup circuit.  Some of the
circuit gadgets used in this section are described in detail
in~\cref{sec:gadgets}.

\subsection{Surface-code specific optimizations and gadgets }

\begin{figure}[t]
    \centering
    \inputtikz{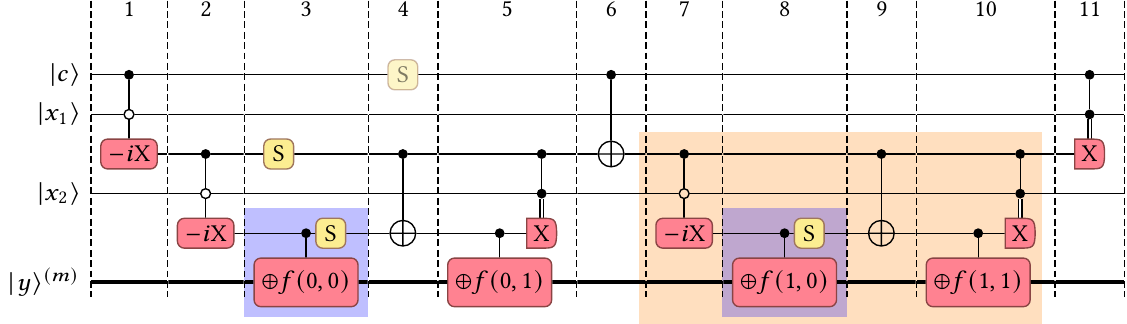}
    \caption[Two-input controlled table lookup circuit with \texttt{AND} gates expanded]{Expanded two-input controlled table lookup circuit. The circuit follows from \eqref{eq:controlled-two-input-table-lookup} and \cref{fig:and-gate}. Register $|c\rangle$ is the control bit, $|x_1\rangle$ and $|x_2\rangle$ are two input bits. We will show that 
    gates in blue areas can be executed simultaneously. The semi-transparent \texttt{S} gate 
    is not present for the two input table lookup, but will be present when the circuit is a part of 
    table lookup with more inputs. Dashed lines indicate layers of gates that are implemented simultaneously.
    The orange area highlights a controlled table lookup circuit with one input.}
    \label{fig:controlled-two-input-table-lookup-and-expanded}
\end{figure}

\begin{figure}[t]
    \centering
    \subfloat[$|S\rangle$ injection via ZZ]{\inputtikz{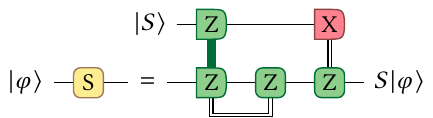}}
    \qquad
    \subfloat[$|S\rangle$ injection via \texttt{CNOT}]{\inputtikz{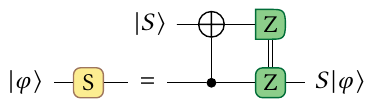}}

    \caption{These circuits apply an \texttt{S} operation to a qubit by consuming and injecting an $|S\rangle$ state.}
    \label{fig:s-injection}
\end{figure}

\begin{figure}[t]
  \centering
  \subfloat[Delayed-choice \texttt{CZ}]{\inputtikz{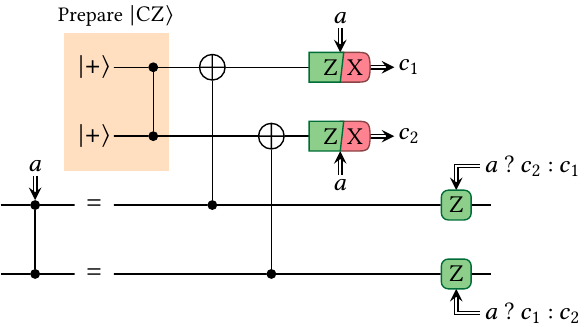}}
  \qquad
  \subfloat[Prepare $|\mathrm{CZ}\rangle$]{\inputtikz{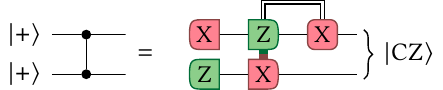}}
  
  \caption{The delayed-choice \texttt{CZ} operation conditionally applies a \texttt{CZ} operation on measurement result $a$ and therefore can be used to implement \texttt{AND}$^\dagger$.  The ``Prepare $|\mathrm{CZ}\rangle$'' operation can be performed before the measurement result is known.}
  \label{fig:delayed-cz}
\end{figure}

\begin{figure}[t]
    \centering
    \inputtikz{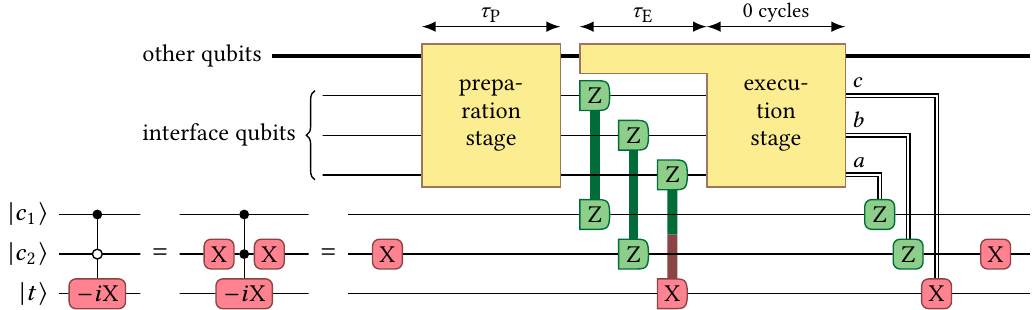}    
    \caption{High-level view of \texttt{CCiX} gadget.
    The \texttt{CCiX} gadget described in \cref{sec:ccix-gadget} consist of three parts: a preparation stage 
    at which $|T\rangle$ states are consumed and $|S\rangle$ states are cloned, three parallel remote joint measurements, and a short execution stage. Clifford operation can be performed on the target qubits right after the remote joint measurements. To apply non-Clifford operations we need to wait for the execution stage to complete and resolve Pauli corrections.}
    \label{fig:ccix-gadget-high-level}
\end{figure}

\begin{figure}[t]
    \centering
    \inputtikz{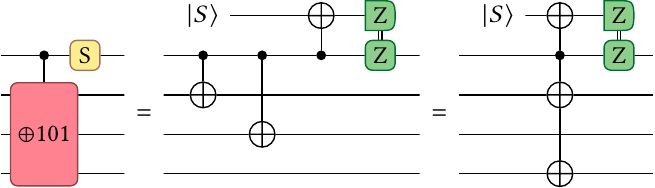}
    \caption{\texttt{CXOR} and $S$ implemented via one multi-target
    \texttt{CNOT}. \texttt{CNOT}s sharing the same control are merged into a
    single multi-target \texttt{CNOT} gates that can be executed in constant
    time independent on the number of targets.}
    \label{fig:cxor-s-via-multi-cx}
\end{figure}

Let us start with an explicit example of a controlled table lookup circuit with
$k=2$ input and $m$ output bits, which follows
from~\cref{fig:controlled-table-lookup}
and~\eqref{eq:controlled-one-input-table-lookup}:
\begin{equation}
    \label{eq:controlled-two-input-table-lookup}
    \inputtikz{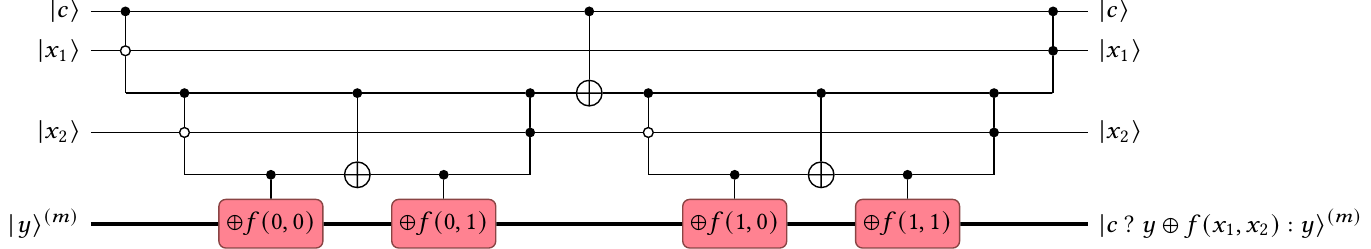}
\end{equation}
We rewrite \texttt{AND} gates in terms of \texttt{CCiX} and $S$ gates as
described in \cref{fig:and-gate} and \texttt{AND}$^{\dagger}$ in terms of $X$
measurement and classically controlled \texttt{CZ} gates. We move the $S$ gates
past the controls of \texttt{CXOR} and \texttt{CCiX} gates as shown in
\cref{fig:controlled-two-input-table-lookup-and-expanded}. Then, we
further replace $S$ gates with the joint measurement-based $|S\rangle$-state
injection circuit shown in \cref{fig:s-injection}(a)\footnote{This step can
likely be simplified by directly implementing $S$ gates in surface code as
described in~\cite{BDM+21}.} and classically-controlled \texttt{CZ} gates with
the delayed-choice \texttt{CZ} gadget~\cite{GF19} depicted in
\cref{fig:delayed-cz}(a). In this gadget, the measurement gates on the top two qubits
measure in the $Z$-basis if the measurement result $a$ is 1, and in the $X$-basis
otherwise.  The ``Prepare $|\mathrm{CZ}\rangle$'' gadget is depicted in
\cref{fig:delayed-cz}(b) and can be executed before the measurement result $a$
is known. We also implement \texttt{CCiX} gates with the fast \texttt{CCiX} gadget
illustrated in \cref{fig:ccix-gadget-high-level} and described
in~\cref{sec:ccix-gadget}. \texttt{CXOR} gates that are followed by $S$ gates
(highlighted by blue rectangles in
\cref{fig:controlled-two-input-table-lookup-and-expanded}) are implemented using
the multi-target \texttt{CNOT} gate as illustrated in
\cref{fig:cxor-s-via-multi-cx} making use of the $|S\rangle$-state injection
circuit from~\cref{fig:s-injection}(b).

\subsection{Logical cycles estimates}
We are ready to determine the number of logical cycles required by the unary-iterate table lookup circuit.  We base our estimates on the duration (number of logical cycles) of the following commonly-used operations:
\begin{equation}
\begin{aligned}
\tau_{\mathrm{RCX}} & : \text{time of remote \texttt{CNOT} gate along an arbitrary path of qubits} \\
\tau_{\mathrm{RZZ}} & : \text{time of remote ZZ measurement along an arbitrary path of qubits} \\
\tau_{\mathrm{RXZ}} & : \text{time of remote XZ measurement along an arbitrary path of qubits} \\
\tau_{\mathrm{CXOR}} & : \text{total time of \texttt{CXOR}} \\
\tau_{\mathrm{CAT}} & : \text{cat state initiation time in \texttt{CXOR} implementation}
\end{aligned}
\end{equation}
We further assume that $X$ and $Z$ measurements as well as $X$ and $Z$
corrections take no time.
We can now discuss in detail the timing of the layers
in~\cref{fig:controlled-two-input-table-lookup-and-expanded}, taking into account
all introduced optimizations.

\paragraph*{Layers 1, 2, and 7 (Total time: $\tau_{\mathrm{RZZ}}$).} The time
for executing the fast \texttt{CCiX} gadget is dominated by the three remote ZZ
and XZ measurements that are shown in~\cref{fig:ccix-gadget-high-level}.  The
preparation stage can always be moved to previous layers if we use two fast
remote \texttt{CCiX} gadgets that we alternate between calls.  For example, the
first \texttt{CCiX} gadget is used in layer 1, while the second one is used in
layer 2, and so on.  Typically XZ measurements take longer than ZZ measurements.
However a small part of the XZ measurement that prepares a Hadamard-conjugated
Bell pair can be done slightly earlier. Consequently, most of the qubits
involved in an XZ or ZZ measurement are occupied for $\tau_{\mathrm{RZZ}}$
logical cycles. To avoid delays due to Pauli corrections that depend on yet
unknown measurement outcomes, we require that the execution stage of the fast
\texttt{CCiX} gadget uses fewer logical cycles than a remote measurement, i.e.,
$\tau_{\mathrm{E}} \le \tau_{\mathrm{RZZ}}$.

\paragraph*{Layer 3 (Total time: $\max(\tau_{\mathrm{RZZ}},
\tau_{\mathrm{CAT}})$).}  A part of the multi-target \texttt{CNOT} gate can be
moved to the next layer, and only the cat state preparation stage needs to be
accounted for in this layer 
(see supplemental material).
At the same time we
apply an $S$ gate via $|S\rangle$-state injection using the circuit
in~\cref{fig:s-injection}(a), which requires $\tau_{\mathrm{RZZ}}$ execution
cycles.  The duration of this layer can therefore be upper bounded by the
maximal duration of these two operations.

\paragraph*{Layers 4, 6, and 9 (Total time: $\tau_{\mathrm{RCX}}$).}  We use the
fastest circuit for $|S\rangle$-state injection and can therefore assume that
the execution time is at most that of a remote \texttt{CNOT} operation.

\paragraph*{Layers 5 and 10 (Total time: $\max(\tau_{\mathrm{CXOR}},
\tau_{\mathrm{RCX}})$).}  We make use of the delayed-choice \texttt{CZ} gate
depicted in~\cref{fig:delayed-cz}.  The $|\mathrm{CZ}\rangle$ preparation can be
executed in previous layers and the two \texttt{CNOT} gates in the
delayed-choice \texttt{CZ} operation can be executed in parallel together with
the \texttt{CXOR} gate.  However, because its control is followed by an $X$
measurement, its outcome must be determined immediately, and therefore we
account for the full execution time of the \texttt{CXOR} gate.

\paragraph{Layer 8: ($\tau_{\mathrm{CAT}}$).}
 This is similar to layer 3, but no $S$ gates are executed.
 
\paragraph{Layer 11: ($\tau_{\mathrm{RCX}}$).}  The cost is dominated by the
\texttt{CNOT} operations in the delayed-choice \texttt{CZ} gate.  Under the
right conditions, this layer can potentially be executed in the same time as the
previous one; however, it is not clear how long of a ladder of dependent
\texttt{AND}$^\dagger$ gates can be executed simultaneously.  We thus opt for a conservative estimate.

\medskip\noindent
Let us define 
\begin{align}
    \tau_R & = \max(\tau_{\mathrm{RCX}},\tau_{\mathrm{RZZ}},\tau_{\mathrm{CAT}}) \\
    \tau_M & = \max(\tau_{\mathrm{RCX}},\tau_{\mathrm{CXOR}}) 
\end{align}

Typically, the cycle times over which we are taking the maximum are equal.  It
is also typically the case that $\tau_{\mathrm{CXOR}} \ge \tau_{\mathrm{RCX}}$.
Therefore we can assume $\tau_R = \tau_{\mathrm{RCX}}$ and $\tau_M =
\tau_{\mathrm{CXOR}}$ in practice. Using this notation, we can bound the number
of logical cycles for a controlled table lookup with $k$ inputs as
$\tau_{\mathrm{CTL}}(k)$.  In the base case, we have
\begin{equation}
    \label{eq:ctl-base-case}
    \tau_{\mathrm{CTL}}(1) \le 3\tau_R + \tau_M,
\end{equation}
where layers 7--9 each contribute $\tau_R$ and layer~10 contributes $\tau_M$ (orange rectangle in~\cref{fig:controlled-two-input-table-lookup-and-expanded}) to the total delay.  From the recursive construction in~\eqref{eq:controlled-one-input-table-lookup}, we derive
\begin{equation}
    \tau_{\mathrm{CTL}}(k) \le 3\tau_R + 2\tau_{\mathrm{CTL}}(k-1), \quad \text{for $k \ge 2$},
\end{equation}
where layers~1, 6, and~11 each incur a delay of $\tau_R$.  Together with~\eqref{eq:ctl-base-case} we can derive the closed form expression
\begin{equation}
    \label{eq:ctl-k}
    \tau_{\mathrm{CTL}}(k) \le 2^{k-1}(6\tau_R + \tau_M) - 3\tau_R, \quad \text{for $k \ge 1$}.
\end{equation}

The number of logical cycles needed for uncontrolled table lookup $\tau_{\text{TL}}(k)$ is
\begin{equation}
    \tau_{\mathrm{TL}}(k) = 2\tau_{\mathrm{CTL}}(k - 1),
    \qquad
    \tau_{\mathrm{TL}}(1) = 2\tau_M.
\end{equation}

We note that these estimates assume that we have enough ancillary qubits to perform all remote operations within any given layer simultaneously, and that a ratio of logical qubits to algorithm qubits of $4:1$ is sufficient for this purpose.

\subsection{Number of logical qubits used}

The number of abstract qubits used by controlled table lookup with $k$ inputs
and $m$ outputs is $2k+m+1$, $2$ qubits per input bit, $1$ qubit per output bit,
and $1$ control qubit. The number of logical qubits needed for table lookup is
\begin{equation}
\label{eq:ctl-qubits}
\sigma_{\mathrm{CTL}}(k) = 4\cdot(2k+1) + O(\sqrt k) + 2\cdot m + O(\sqrt m) + O(1)
\end{equation}
The constant term is the number of qubits required for the \texttt{CCiX} and
delayed-choice \texttt{CZ} gadgets, qubits for $|S\rangle$-state delivery and
some additional padding qubits to fit the input qubits in a square shape. The
number of output qubits is multiplied by $2$ corresponding to the number of
logical qubits per target in the \texttt{CXOR} gadget as explained in
supplemental material.
We assume that the target qubits are aligned in a
square shape and use a dedicated routine to create a cat state in every second
column of that square following a snake like path.  This incurs on overhead of
at most $O(\sqrt m)$ additional qubits.  The number of the other abstract qubits
(control qubit, input qubits, and helper qubits) is multiplied by four to ensure
that all layers described in the previous subsection can be executed as fast as
possible.  To ensure that we can route the remote measurements efficiently from
these input qubits to the \texttt{CCiX} gadgets we align them in a squarish
shape and also use the qubits at the boundary of that shape as auxiliary qubits,
which results in the $O(\sqrt k)$ term in~\eqref{eq:ctl-qubits}.

The number of \texttt{CCiX} gadgets and delayed-choice \texttt{CZ} gadgets is
constant and is calculated based on duration of the gadgets and remote ZZ
measurement. We need to make sure that consecutive dependent \texttt{CCiX} or
delayed-choice \texttt{CZ} gates  can be executed without delays (as, e.g., in
layers 1 and 2 in \cref{fig:controlled-two-input-table-lookup-and-expanded}).
The number of qubits required for uncontrolled table lookup on $k$ qubits is the
same as the number of qubits for controlled table lookup on $k-1$ qubits, since
we use the control qubit for the most-significant bit:
\begin{equation}
\label{eq:tl-qubits}
\sigma_{\mathrm{TL}}(k) = \sigma_{\mathrm{CTL}}(k - 1)
\end{equation}


\paragraph*{Layout.}

\newcommand{\layoutGrid}{%
    \checkerboard{17}{22}
    
    
    \begin{scope}[on background layer]
    \begin{scope}
    \clip (-.5,-.5) rectangle ++(2,9);
    \node[ft h,minimum width=1.45cm,minimum height=4.95cm] at (0,3.5) {\rotatebox{90}{Multi-target \texttt{CNOT}}};
    \end{scope}
    \node[ft h,minimum width=2.95cm,minimum height=4.45cm] at (12.5,4) {\texttt{CCiX}};
    \node[ft h,minimum width=2.95cm,minimum height=4.45cm] at (18.5,4) {\texttt{CCiX}};
    
    \path[fill=blue,fill opacity=.25,draw=black] (9.5,14.5) rectangle ++(2,2);
    \path[fill=fill-h,fill opacity=.25,draw=black] (9.5,8.5) rectangle ++(12,6);
    \path[fill=fill-bell,fill opacity=.25,draw=black] (12.5,8.5) rectangle ++(3,3) (18.5,8.5) rectangle ++(3,3);
    \path[fill=fill-z,fill opacity=.25,draw=black]
        (1.5,-.5) -| ++(2,9) -| ++(-1,-8) -| cycle;
    \path[fill=orange,fill opacity=.15] (-.5,8.5) -| ++(10,8) -| ++(-2,-6) -| cycle (-.5,10.5) |- ++(8,6) |- ++(-6,-2) |- cycle;
    \draw (-.5,8.5) rectangle ++(10,8);
    \path[fill=orange,fill opacity=.3,draw=black] (1.5,10.5) rectangle ++(6,4);
    \end{scope}
    
    \foreach \x/\y in {0/9,2/9,4/9,6/9,8/9,0/11,8/11,0/13,8/13,0/15,2/15,4/15,6/15,8/15} {%
      \fill (\x,\y) circle (2pt);
    }
    \fill[draw-x] (1,8) circle (2pt);

    \begin{scope}[every node/.style={font=\scriptsize}]
      \node at (13,8) {$|c_2\rangle$};
      \node at (14,8) {$|c_1\rangle$};
      \node at (15,8) {$|t\rangle$};

      \node at (19,8) {$|c_2\rangle$};
      \node at (20,8) {$|c_1\rangle$};
      \node at (21,8) {$|t\rangle$};
    \end{scope}
}

\begin{figure}[t]
    \centering
    \inputtikz{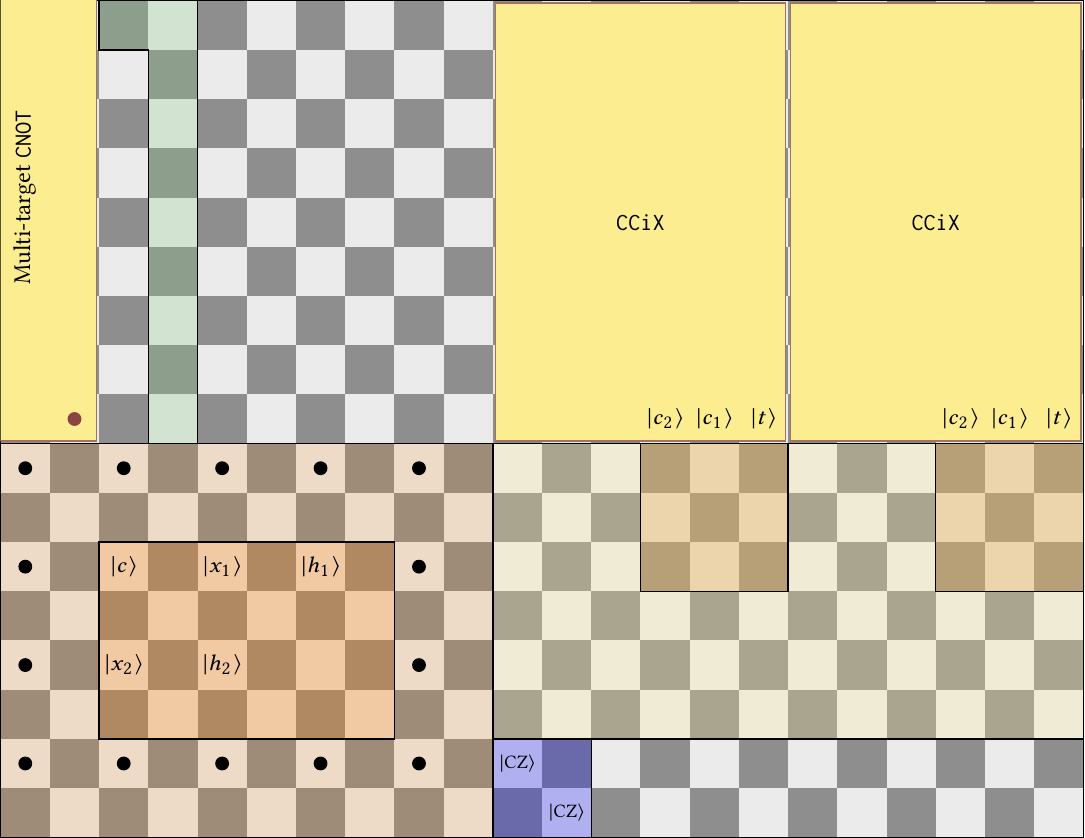}
    \caption{Arrangement of input qubits, output qubits and gadgets for a controlled 2-input table lookup instance with 2 \texttt{CCiX} gadgets.}
    \label{fig:layout-ctl2}
\end{figure}

\cref{fig:layout-ctl2} provides an overview of the layout, illustrated for a
controlled table lookup instance using 2 input bits, 2~\texttt{CCiX} gadgets,
and 1~delayed-choice \texttt{CZ} gate. For the $2k + 1$ input qubits and control
qubit we use a 4:1 ratio of auxiliary qubits, i.e., for each data qubit there
are 3 auxiliary qubits.  These are placed in patches of $2\times2$
qubits~\scalebox{.4}{\inputtikz{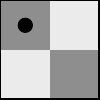}}, where each data qubit is in
the top-left corner.  All these patches are aligned in a square shape region and
also the boundary around this square shape of $2\times2$ patches with 1 data
qubits and 3 auxiliary qubits is reserved for routing.  This region is
highlighted in orange in~\cref{fig:layout-ctl2} (dark orange for the input
qubits and a bit lighter orange for the boundary).  The black dots in the
boundary are assumed to be some data qubits which may be in a non-blank state
but do not require their auxiliary qubits during the execution of the table
lookup operation.

The output qubits are part of the \texttt{CXOR} gadget in the top-left corner.
The dot in the bottom right is the interface qubit to establish the remote ZZ
measurement to initiate the fanout as described in
supplemental material.
We also assume that all output qubits are arranged in a square with the cat state
arranged in a snake pattern.  Some auxiliary qubits are required for the holes in the
cat state at the corner of the snake pattern.  However, these won't exceed
$4\sqrt{m}$ qubits.

The two~\texttt{CCiX} gadgets are located at the top-right
in~\cref{fig:layout-ctl2}.  Each spans $9\times 6$ qubits, as described
in~\cref{sec:ccix-gadget}.  Below the \texttt{CCiX} gadgets and right of the
input qubits, there is an area of $6\times 12$ qubits, highlighted in yellow,
that is used to connect input qubits to the interface qubits of the
\texttt{CCiX} gadgets (dots in the lower left corner) via joint ZZ and XZ
measurements and teleportation circuits.  We will always route these remote
measurements through the top~3 qubit patches in the right boundary of the input
qubit region.  In order to account for permutations of the target and control
qubits in the \texttt{CCiX} gadgets, we make use of a $3\times 3$ switch board
below each \texttt{CCiX} gadget as described
in~\cref{sec:switch-board}.

The $9$ qubits highlighted in green above the input qubit region is to deliver
$|S\rangle$-states into the input qubit region.  Finally, the $2\times 2$ qubits
highlighted in blue will contain $|\mathrm{CZ}\rangle$ states to execute
delayed-choice \texttt{CZ} gates.  Following this layout, we can
state~\eqref{eq:ctl-qubits} more precisely as
\begin{equation}
\sigma_{\mathrm{CTL}}(k) = 4(c + 2)(r + 2) + 90 \cdot \#\mathtt{CCiX} + 4 \cdot \#\mathtt{CZ} + 2m + O(\sqrt m),
\end{equation}
where $c = \lceil\sqrt{2k+1}\rceil$ and $r =
\left\lceil\frac{2k+1}{c}\right\rceil$ are the number of columns and rows to fit
the $2k+1$ input qubits into a square shape, and $\#\mathtt{CCiX}$ and
$\#\mathtt{CZ}$ are the number of \texttt{CCiX} and \texttt{CZ} gadgets,
respectively.

\paragraph*{Layer execution.}  Next, we describe how the operations in each
layer in~\cref{fig:controlled-two-input-table-lookup-and-expanded} are executed
using the layout in~\cref{fig:layout-ctl2}.  The first two layers execute the
\texttt{CCiX} gates.  Recall that the preparation stages of the \texttt{CCiX}
gates can be done in previous layers.
\[
\label{eq:execute-1-2}
\scalebox{.6}{\inputtikz{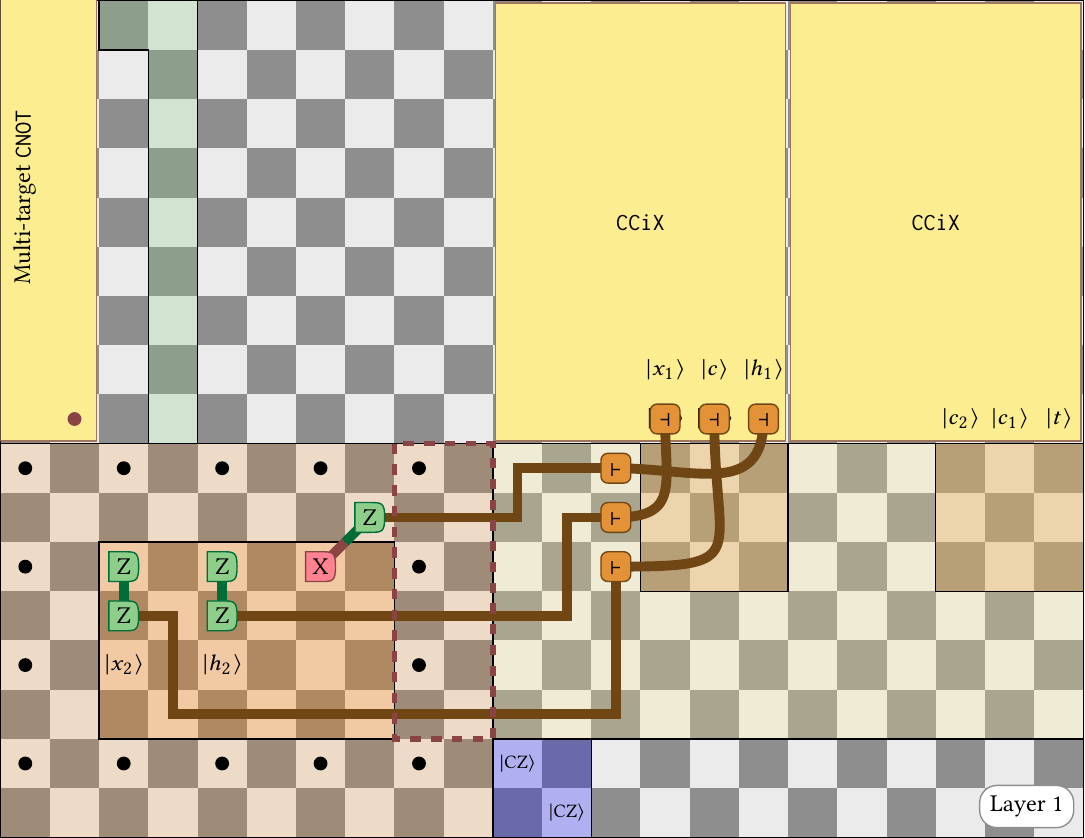}}
\quad
\scalebox{.6}{\inputtikz{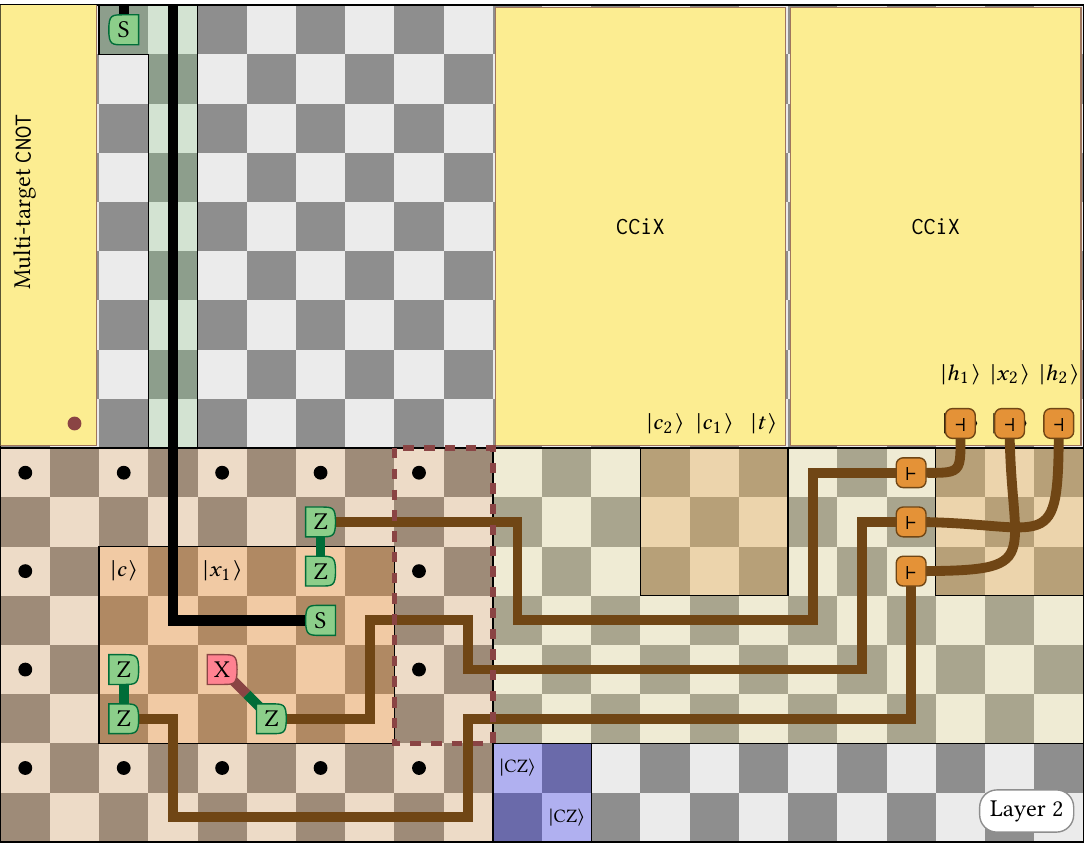}}
\]
During the preparation stage of the \texttt{CCiX} gates, also the teleportation
circuits in the switch boards are performed (see~\eqref{eq:switch-boards}).  We
then execute the remote joint measurements using more teleportation circuits
using remote Measure ZZ \& Move operation
(see~\cref{sec:gadgets})
together with joint ZZ and XZ measurements
for the control lines and the target line, respectively.  The brown lines
between the teleportation targets in the switch board and the joint measurements
are also teleportation circuits.  The joint measurements can be executed at the
same time as these. We alternate between the \texttt{CCiX} gadgets such that the
preparation stages do not interfere. The routing of the teleportation paths
between the joint measurements and the switch board is done in a way that the
paths always exit in the top three $2\times 2$ qubit patches of the boundary
(the dashed rectangle).  The measurement in the top patch connects to the
top-most qubit in the switch board, whereas the measurement in the bottom patch
connects to the bottom-most qubit in the switch board.  To route them in
parallel, we need three rows of qubit patches as in Layer~1, or four rows of
qubit patches if additionally an $|S\rangle$-state needs to be delivered as in
Layer~2 for the injection circuit in the next layer. Because there are 6 rows of
helper qubits below the \texttt{CCiX} gadgets, teleportation paths can be routed
along switch board gadgets when \texttt{CCiX} gadgets further to the right are
addressed, as shown in Layer~2.

Layer~3 will use two $|S\rangle$-states that are prepared in Layer~2. One is for
the $S$ gate on~$|h_2\rangle$, which is executed together with the \texttt{CXOR}
gate as described in~\cref{fig:cxor-s-via-multi-cx}, and the other one on qubit
$|h_1\rangle$, which is prepared below it, such that we can use the $|S\rangle$
injection circuit that uses joint ZZ measurement (see
\cref{fig:s-injection}(a)).  In Layer~3, we execute the \texttt{CXOR} gate by
performing a remote joint ZZ measurement with its interface qubit.  The
$|S\rangle$ is consumed to apply an $S$ gate on $|h_1\rangle$ and optionally
another $|S\rangle$-state is delivered right above $|c\rangle$ in case it needs
to be applied in Layer~4.
\begin{equation}
\label{eq:execute-3-4}
\scalebox{.6}{\inputtikz{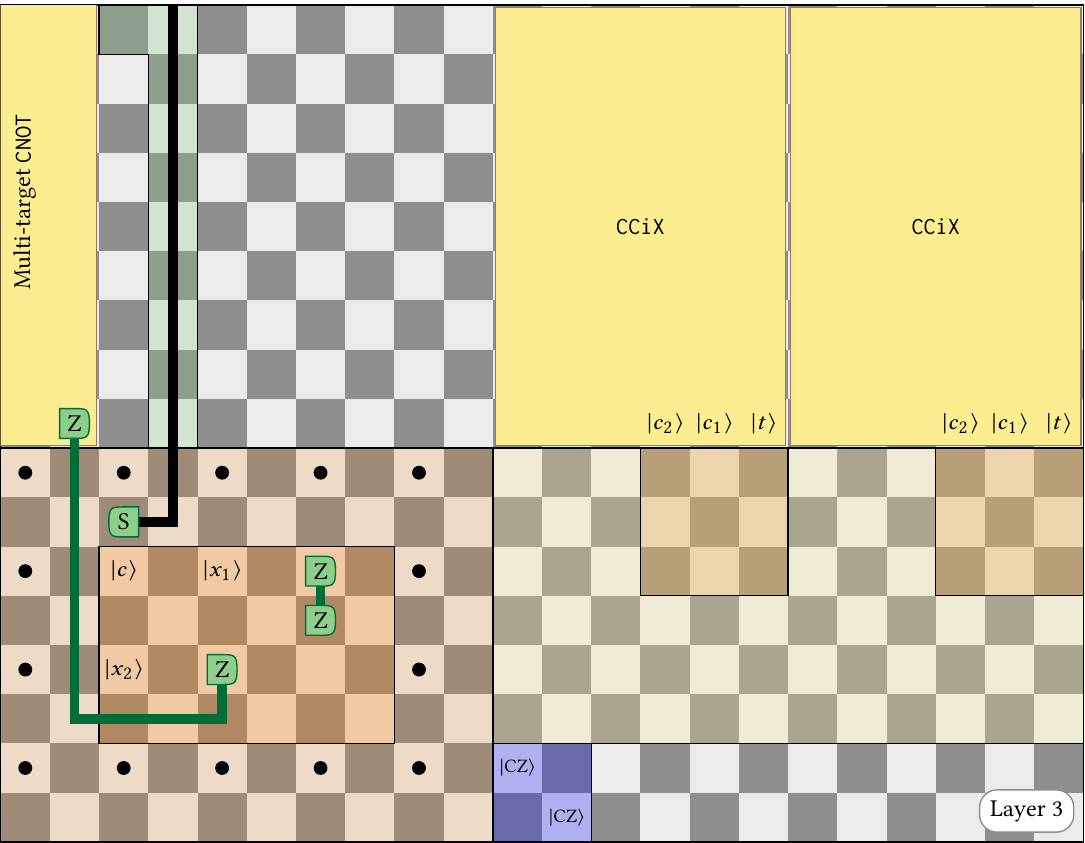}}
\quad
\scalebox{.6}{\inputtikz{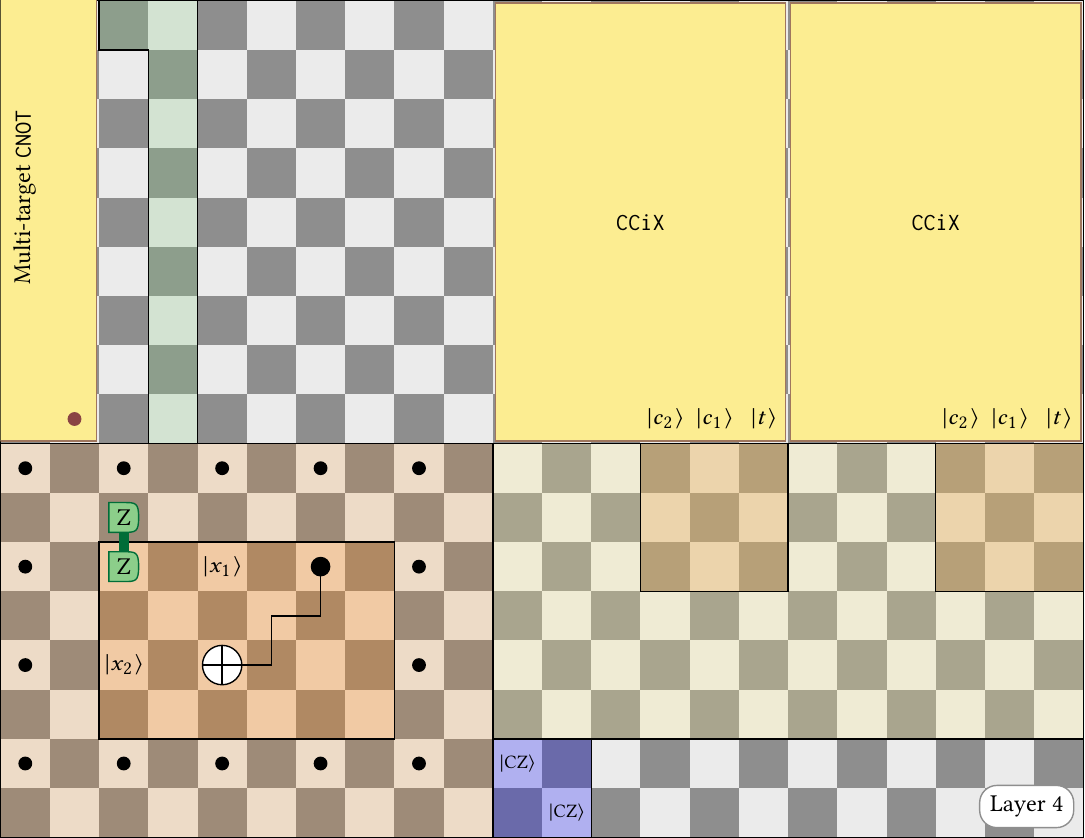}}
\end{equation}
In Layer~4, if the $S$ gate is applied, it injects the $|S\rangle$-state via
joint ZZ measurement; it also executes a remote \texttt{CNOT} gate controlled on
$|h_1\rangle$ targeted on $|h_2\rangle$. Due to the $2\times 2$ patches, a route
can readily be found.

Layer~5 executes another \texttt{CXOR} gate as in Layer~3 and also connects the
two control qubits $|h_1\rangle$ and $|h_2\rangle$ of the \texttt{AND}$^\dagger$
operation to the $|\mathrm{CZ}\rangle$ states via two remote \texttt{CNOT}
operations to perform the delayed-choice \texttt{CZ} gate.  These are routed
such that they exit the input area via the topmost patch and the fourth patch
from the top at the right boundary.  Recall that the $|\mathrm{CZ}\rangle$
states are prepared in previous layers.  Layers~6--10 are similar to Layers~1--5
and Layer~11 is another example of a delayed-choice \texttt{CZ} gate.
\[
\label{eq:execute-5-11}
\scalebox{.6}{\inputtikz{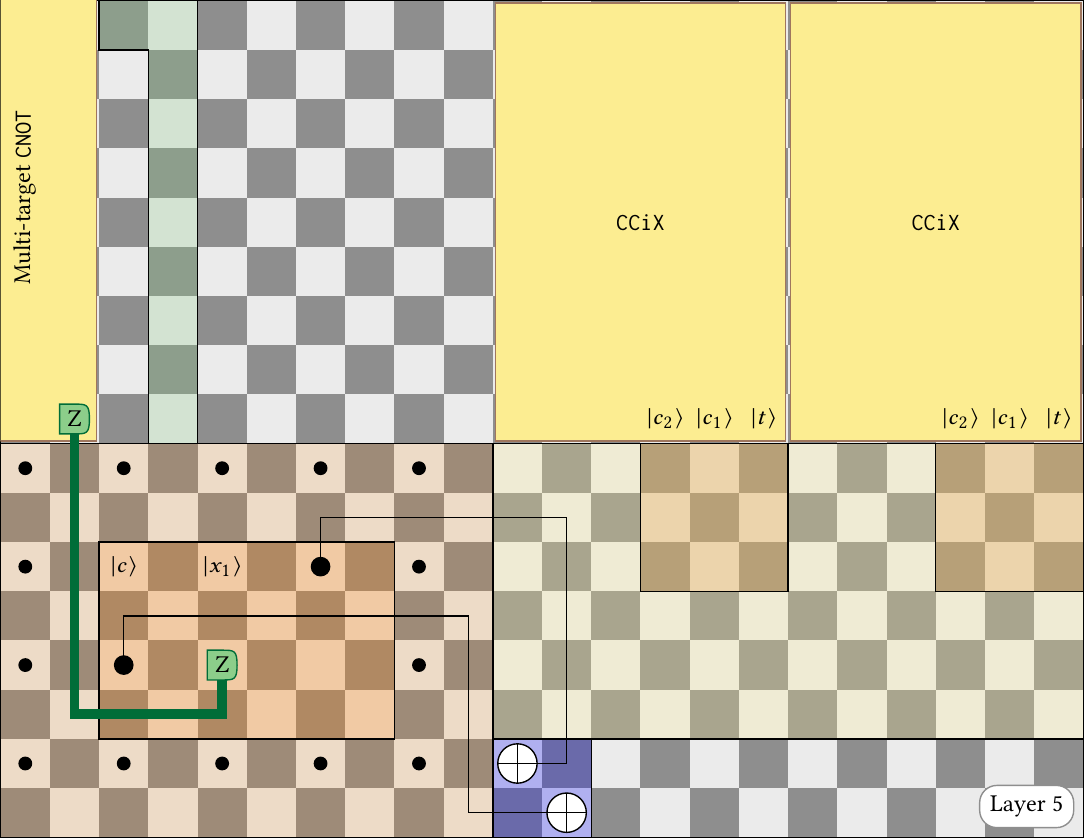}}
\quad
\scalebox{.6}{\inputtikz{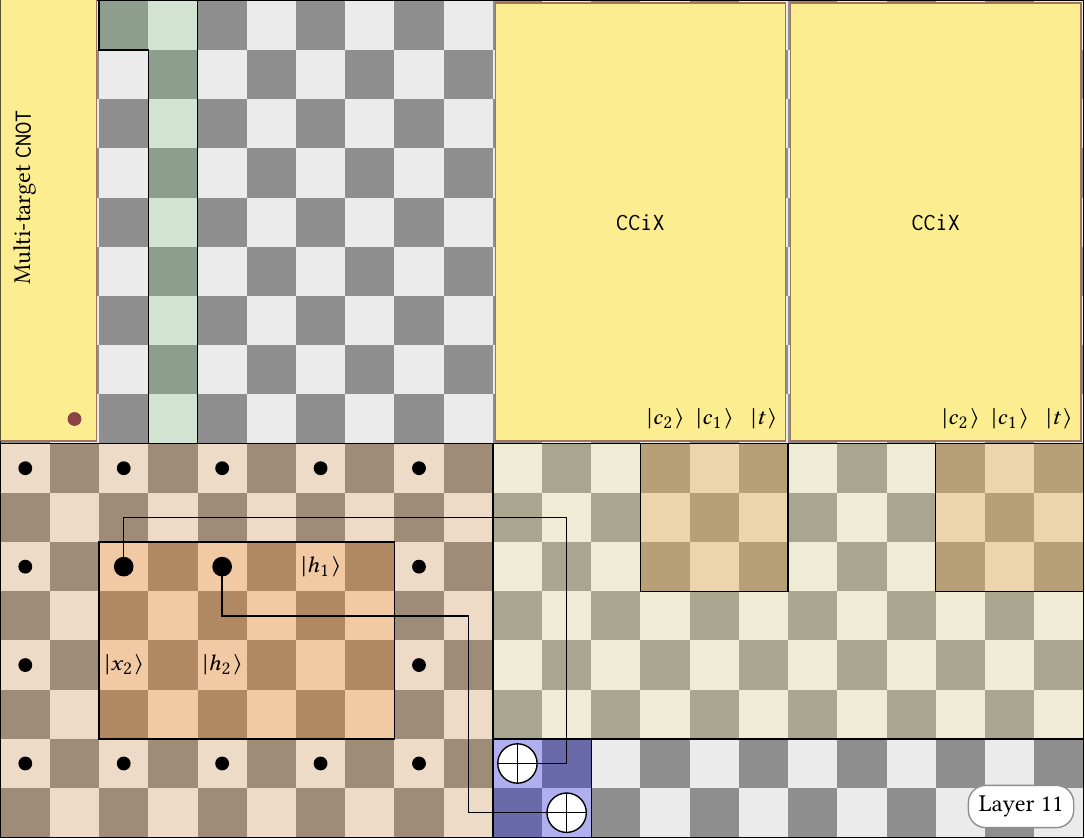}}
\]

\subsection{T state consumption rate}
The highest consumption rate of $|T\rangle$-states happens during the execution
of consecutive  \texttt{CCiX} gates (as, e.g., in the first two layers in
\cref{fig:controlled-two-input-table-lookup-and-expanded}). It is reasonable to
average out the consumption rate over a small window of logical cycles using small $T$-state
buffers. The smoothed-out peak number of required $|T\rangle$-states per
cycle is: 
\begin{equation}
    \frac{\text{Number of T states per CC-iX}}{\tau_{\mathrm{RZZ}}} = \frac{4}{\tau_{\mathrm{RZZ}}}
\end{equation}
When running several instances of this circuit in parallel we simply multiply
the required rate by the number of circuits running in parallel.


\section{Space-time trade-offs}
\label{sec:tradeoff}
In this section, we describe optimizations that allow for space-time trade-offs.
The zipper construction copies all input qubits $p>1$ times using the fanout gadget and
performs $p$ table lookups simultaneously. Depending on the number of output qubits and $p$,
it may make sense to use separate output registers for the $p$ simultaneous table lookups,
merging the outputs into a single register only once both lookups have completed (using \texttt{CXOR} gates).
In the case of a large number of output qubits, the two table lookups may write into
the same output register by interleaving the execution of the \texttt{CXOR} gates
that write to the output register.
In the following description, we choose the latter approach and $p=2$.

\subsection{Zipper construction}
\label{sec:zipper}
\begin{figure}[t]
    \centering
    \footnotesize
    \subfloat[\label{fig:controlled-table-lookup-zipper}]{\inputtikz{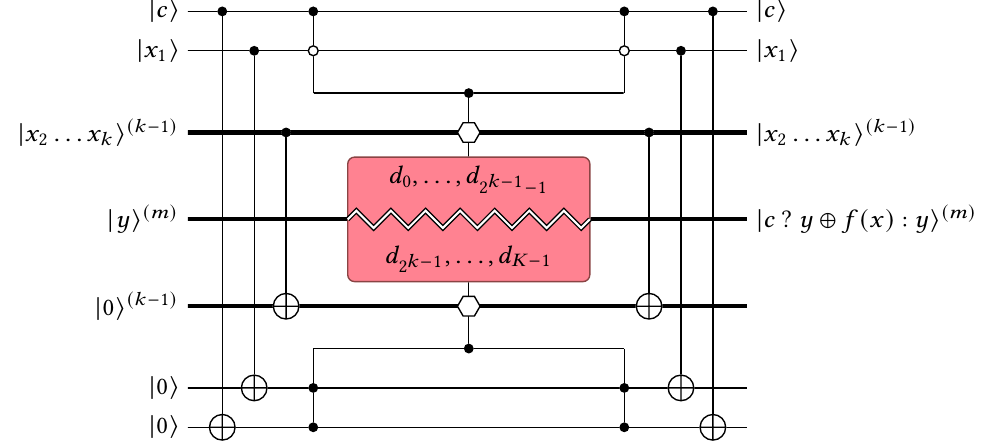}{circuits/ctl-tl-zipper}}

    \subfloat[\label{fig:unary-iterate-table-lookup-zipper}]{\inputtikz{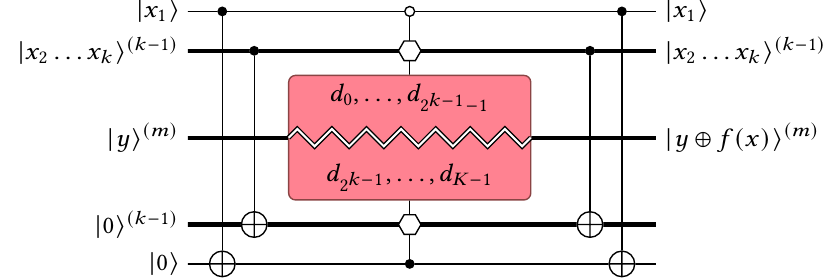}{circuits/nctl-tl-zipper}}
    \caption{parallel (a) controlled and (b) uncontrolled $k$-input and $m$-output table lookup circuit; we use a hexagon to denote the input register for the table lookup.  The idea is to interleave the \texttt{CXOR} gates in each table lookup instance}
    \label{fig:controlled-table-lookup-zipper-tmp}
\end{figure}
\cref{fig:controlled-table-lookup-zipper-tmp} illustrates the zipper
construction for both controlled and uncontrolled table lookup on $k$ input bits
and $K$ entries.  Fanout gadgets copy all input qubits.  Then the upper part
writes entries $d_x$ for which the most-significant bit $x_1 = 0$ in the index
$x$, whereas the lower part writes entries for which $x_1 = 1$.  Both table
lookups share the same target qubit register $|y\rangle$ but no other qubits.
The idea is to alternate and interleave the executions of \texttt{CXOR} gates,
while executing the other operations in parallel as they act on independent
qubits.  The controlled table lookup in~\cref{fig:controlled-table-lookup-zipper}
and the uncontrolled table lookup in~\cref{fig:unary-iterate-table-lookup-zipper}
are very similar except for the copying of input qubits.  Also note that the
controlled zipper construction requires one more \texttt{AND} gate compared to
the standard construction in~\cref{fig:controlled-table-lookup}.

\begin{figure}[t]
    \centering
    \inputtikz{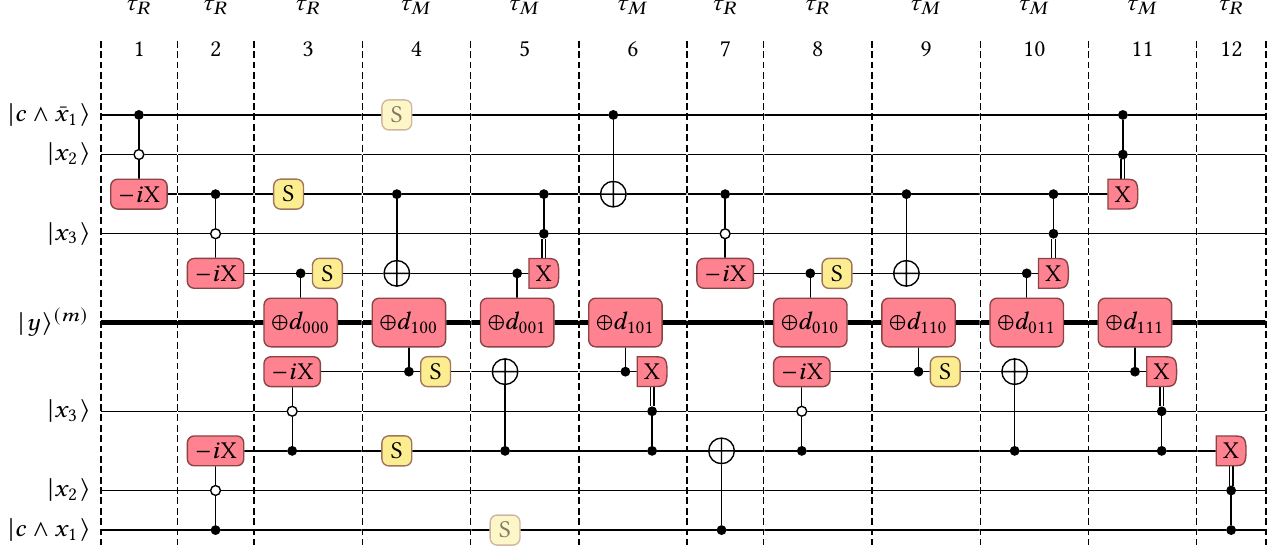}{circuits/ctl-tl-zipper-expanded}
    \caption{Similar
    to~\cref{fig:controlled-two-input-table-lookup-and-expanded}, we show an
    expansion of operations assigned to different layers.  Here two instances
    are executed in parallel using the zipper construction.  Note how the lower
    instance is shifted by one layer and how the \texttt{CXOR} gates alternate.}
    \label{fig:ctl-tl-zipper-expanded}
\end{figure}
\cref{fig:ctl-tl-zipper-expanded} illustrates the interleaving when $k=3$, split
on $x_1$ and data lookups for $x_2$ and $x_3$ interleaved.  Note that the
construction requires one more layer
than~\cref{fig:controlled-two-input-table-lookup-and-expanded} because the lower
table lookup is shifted by one layer.  Furthermore, the execution times for layers
change.  Some layers now require $\tau_M$ instead of $\tau_R$ since we expect
that the execution of \texttt{CXOR} takes longer than a remote \texttt{CNOT}.

From layers 7--11, we can derive that a controlled table lookup with the zipper
construction takes
\begin{equation}
    \hat\tau_{\mathrm{CTLz}}(2) = 2\tau_R + 3\tau_M
\end{equation}
cycles.  This time does not take into account the copying of input qubits and
the additional \texttt{AND} and \texttt{AND}$^\dagger$ gates
in~\cref{fig:controlled-table-lookup-zipper}.  Together with layer 1, 6, and 12,
we derive the recursion
\begin{equation}
    \hat\tau_{\mathrm{CTLz}}(k) = 2\tau_R + \tau_M + 2\hat\tau_{\mathrm{CTLz}}(k - 1) \qquad \text{for $k \ge 3$},
\end{equation}
which can be expressed in closed form as
\begin{equation}
    \hat\tau_{\mathrm{CTLz}}(k) = 2^k(\tau_R + \tau_M) - 2\tau_R - \tau_M \qquad \text{for $k \ge 2$}.
\end{equation}
Finally, taking into account the overhead from copying input qubits and
additional \texttt{AND} and \texttt{AND}$^\dagger$ gates the cost is
\begin{equation}
    \tau_{\mathrm{CTLz}}(k) = 2^k(\tau_R + \tau_M) - \tau_M + O(\sqrt k) \qquad \text{for $k \ge 2$}.
\end{equation}
Here, we account for an $O(\sqrt k)$ overhead of copying input qubits assuming
that they are aligned in a square, as described in the previous section.

\subsection{Experimental results}
\begin{table}[t]
\caption{Comparing expected logical cycles for controlled table
lookup~\eqref{eq:ctl-k} with $m=2$, $\#\mathtt{CCiX} = \min\{k, 2\}$,
$\mathtt{CZ} = 2$, and $m = 7$ to numbers obtained from an implementation
($\tau_R = 5$ and $\tau_M = 7$).}
\label{tab:experimental-results}
\centering
\begin{tabularx}{.6\linewidth}{Xrrrr}
    \toprule
    $k$ & $\sigma_{\mathrm{CTL}}(k)$ & $\tau_{\mathrm{CTL}}(k)$ & $\tau'_{\mathrm{CTL}}(k)$ & $\tau_{\mathrm{CTLz}}(k)$ \\
    \midrule
    $1$ & 186 &    22 &    22 &   --- \\
    $2$ & 292 &    59 &    44 &    41 \\
    $3$ & 312 &   133 &    90 &    89 \\
    $4$ & 312 &   281 &   214 &   185 \\
    $5$ & 332 &   577 &   455 &   377 \\
    $6$ & 356 & 1,169 &   805 &   761 \\
    $7$ & 356 & 2,353 & 1,812 & 1,529 \\
    $8$ & 380 & 4,721 & 3,229 & 3,065 \\
    \bottomrule
\end{tabularx}
\end{table}

We have implemented the layout algorithm for a controlled table lookup using the
standard construction on $k$ inputs and $K = 2^k$ entries with $\#\mathtt{CCiX}
= \min \{2, k\}$ and $\#\mathtt{CZ} = 1$.  We choose $m=7$ and used a simple
vertical cat state layout in the implementation of the \texttt{CXOR} gadget.
\cref{tab:experimental-results} lists the results, assuming that preparing
$|0\rangle$ and $|+\rangle$ states requires 1~logical cycle, preparing
$|S\rangle$ and $|T\rangle$ states requires 5~logical cycles, joint XX and ZZ
measurement takes 2~logical cycles, and joint XZ measurement takes 3~logical
cycles. \todo{why these numbers? reference?} Therefore, $\tau_R = 5$ and $\tau_M = 7$.  In the implementation there
are no strict barriers between the layers which can reduce the effective overall
runtime, denoted by $\tau'_{\mathrm{CTL}}(k)$, by up to 1/3 compared to the upper
bound.  We also list the upper bound for the zipper construction
$\tau_{\mathrm{CTLz}}(k)$ for $k \ge 2$.  The initialization time, consisting of
$|S\rangle$-state preparation for the \texttt{CCiX} gadgets and the initial
preparation stage of the first \texttt{CCiX} execution, is 20 logical cycles and
is not included in any of the estimates.  Also the initialization time for
copying qubits is not included in $\tau_{\mathrm{CTLz}}(k)$.

\section{Commonly used gadgets}
\label{sec:gadgets}

\begin{table}[t]
\caption{\label{tab:extended-operatrions}
Extended set of surface code operations that can be performed in constant depth. 
We provide contant depth circuits for all of these operaions in the supplementary material.}
\begin{tabular}{|c|c|}
\hline 
Operation type & Operation Name(s)\tabularnewline
\hline 
\hline 
\textbf{Nearest neighbour} & Move (\cref{fig:axis-independent-operations-main})\tabularnewline
\cline{2-2} 
adjacent  & Prepare Bell (\cref{fig:axis-independent-operations-main})\tabularnewline
\cline{2-2} 
target qubits & Measure Bell (\cref{fig:axis-independent-operations-main})\tabularnewline
\hline 
\hline 
\textbf{Remote two-qubit} & Measure ZZ, Measure XX\tabularnewline
\cline{2-2} 
control \& target qubits  & Measure XX \& Move, Measure ZZ \& Move\tabularnewline
\cline{2-2} 
connected by a path & Measure XZ\tabularnewline
\cline{2-2} 
 & \texttt{CNOT}\tabularnewline
\cline{2-2} 
 & Prepare Bell, Measure Bell\tabularnewline
\cline{2-2} 
 & Move\tabularnewline
\hline 
\hline 
\textbf{Remote multi-qubit} & Prepare x-cat state (\cref{fig:cat-state-main}) \tabularnewline
\cline{2-2} 
   & Prepare z-cat state (\cref{fig:cat-state-main})\tabularnewline
\cline{2-2} 
& Fanout (\cref{eq:fanout-main}) \tabularnewline
\cline{2-2} 
& Multi-target \texttt{CNOT}\tabularnewline
\cline{2-2} 
 & \texttt{CXOR} (\cref{fig:cxor})\tabularnewline
\hline 
\end{tabular}
\end{table}

We first review an extended set of surface code operations \cref{tab:extended-operatrions}, which can be implemented as constant depth circuits 
using a minimal set of surface code operations in \cref{fig:surface-code-operations} as explained in this section and supplementary material.
Alternatively, many of these operations can be implemented directly in the surface code in a more optimized way. 

\begin{figure}
    \centering
    \inputtikz{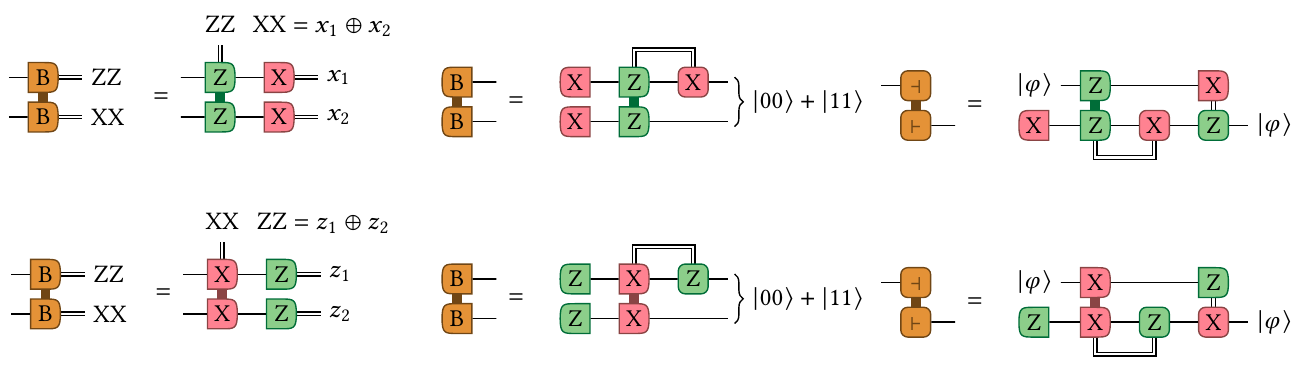}
    \caption[Axis independent operations]{
    Axis independent Bell measurement, Bell state preparation and Move operations.
    }
    \label{fig:axis-independent-operations-main}
\end{figure}

It is tedious to always have to deal with horizontal and vertical directions for XX and ZZ measurements, 
so the first step is to introduce axis-independent gadgets of commonly-used operations. In~\cref{fig:axis-independent-operations-main}, we present axis-independent circuits for Bell measurement, Bell preparation, and the move operation, which can be used to move the state of a qubit to an ancillary qubit.
Their remote versions also have no axis related geometric restrictions.

Remote operations inherit restrictions of their nearest-neighbor versions. 
Target qubits for joint measurements must be aligned according to the axis. 
For example, target qubits of remote ZZ measurement must be vertically adjacent to the path of helper qubits connecting the target qubits.
Similarly, control qubit of CNOT, CXOR and Multi-target CNOT must be vertically adjacent to the path of helper qubits; 
all the target qubits must be horizontally adjacent to the path of helper qubits.
Measure and Move operations inherit have the same geometric restrictions as applying measure and move sequentially, but have a smaller depth. 
Remote measure $XZ$ requires a square of qubits along the path mimicking restriction of the local version.

State $z$-cat can be prepared on a set of vertical segments of a path.
Similarly, state $x$-cat can be prepared on a set of horizontal segments of a path.

The quantun fanout operation achieves the mapping
\begin{equation}
    \label{eq:fanout-main}
    (\alpha\ket{0}+\beta\ket{1}) \otimes |0^{n-1}\rangle \mapsto
    \alpha|0^n\rangle + \beta|1^n\rangle.
\end{equation}
This operation can be implemented in constant depth by first creating a $z$-cat state on the latter $n-1$ qubits, which results in
\[
  \tfrac{\alpha}{\sqrt2}|0^n\rangle + \tfrac{\beta}{\sqrt2}|1^n\rangle
  + \tfrac{\alpha}{\sqrt2}|01^{n-1}\rangle + \tfrac{\beta}{\sqrt2}|10^{n-1}\rangle.
\]
By applying a joint ZZ measurement on the first two qubits, one either obtains the expected result when the measurement outcome corresponds to the +1 eigenvalue or, otherwise, an $X$-correction on each of the last $n-1$ qubits is needed.

\begin{figure}[t]
    \caption{Cat state preparation on 2 or more vertically adjacent qubits.  The cat state preparations used on the right-hand sides are applied to an even number of qubits.  Further, consecutive Pauli corrections can be merged by computing the parity of respective measurement results.}
    \label{fig:cat-state-main}
    \footnotesize
    \medskip
    
    \subfloat[$n=2$]{\scalebox{.9}{\inputtikz{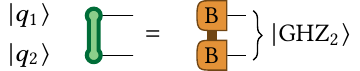}}}
    \hfill
    \subfloat[$n>2$, $n$ odd]{\scalebox{.9}{\inputtikz{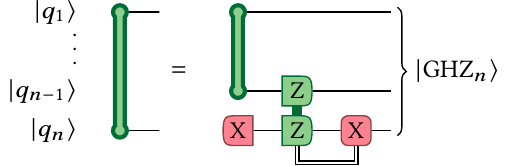}}}
    \hfill
    \subfloat[$n>2$, $n$ even]{\scalebox{.9}{\inputtikz{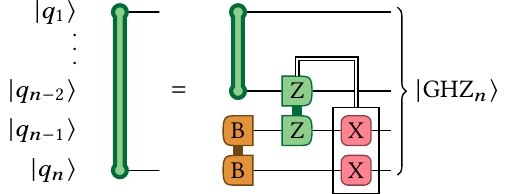}}}
    
    \hrulefill
    \medskip
    
    \subfloat[$n=2$]{\scalebox{.9}{\inputtikz{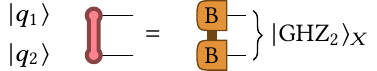}}}
    \hfill
    \subfloat[$n>2$, $n$ odd]{\scalebox{.9}{\inputtikz{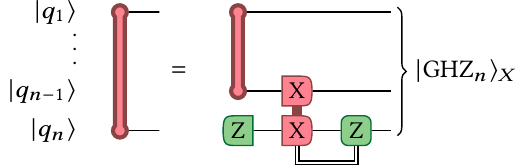}}}
    \hfill
    \subfloat[$n>2$, $n$ even]{\scalebox{.9}{\inputtikz{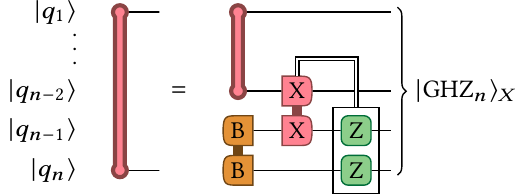}}}
\end{figure}

\subsection{Switch board construction.}
\label{sec:switch-board}

For multiple parallel Bell preparations or teleportations of $k$ qubit pairs
$(q_1, q'_{\pi(1)}), \dots, (q_k, q'_{\pi(k)})$ with some arbitrary permutation
$\pi \in S_k$, we make use of the following switch board gadgets, expanding on the idea in \cite{OptimalCircuitsForNNA}:
\begin{equation}
\label{eq:switch-boards}
\inputtikz{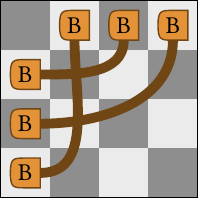}
=
\inputtikz{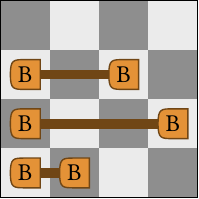}
+
\inputtikz{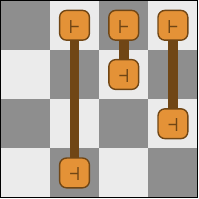}
\qquad
\inputtikz{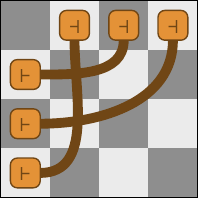}
=
\inputtikz{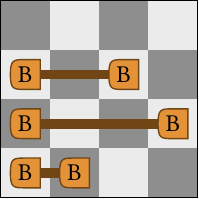}
+
\inputtikz{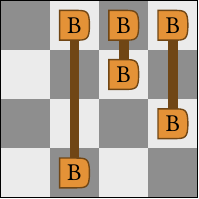}
\end{equation}
Source qubits are aligned horizontally in adjacent columns above the switch
board and target qubits are aligned vertically in adjacent rows to the left of the
switch board.  Depending on whether to prepare Bell states or teleport states,
one first prepares remote Bell pairs with target qubits in each row targeting
the respective column, and then teleport circuits (for Bell pair preparation) or
remote Bell measurement (or teleportation) in each column.

\subsection{CCiX gadget}
\label{sec:ccix-gadget}
\begin{figure}[t]
\centering
\subfloat[]{\scalebox{.85}{\inputtikz{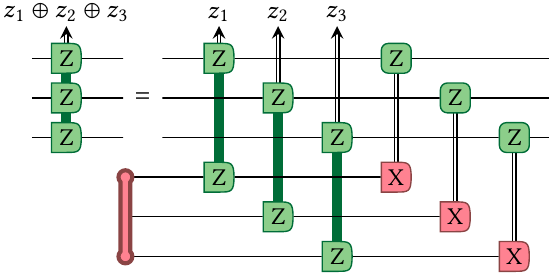}}}
\qquad\qquad
\subfloat[]{\scalebox{.85}{\inputtikz{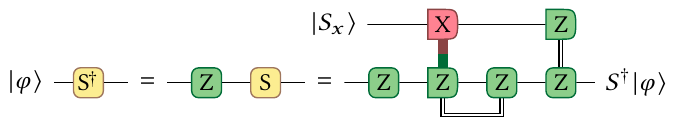}}}
\caption{(a) Multi-qubit joint ZZ measurement using an $x$-cat state and joint ZZ measurement; (b) $S^\dagger$ gate via $|S_x\rangle$ injection with XZ measurement}
\label{fig:multi-qubit-zz-measurement}
\end{figure}

In this section we derive a fast version of the \texttt{CCiX} gadget defined in~\cref{fig:and-gate} with its implementation outlined in~\cref{fig:ccix-gadget-high-level}. 
This is an alternative to using AutoCCZ state introduced in \cite{AutoCCZ}.
We start the derivation by decomposing the double-controlled $-\mathrm{i}Z$ gate into four exponentials of multi-qubit Pauli operators:
\begin{equation}
\inputtikz{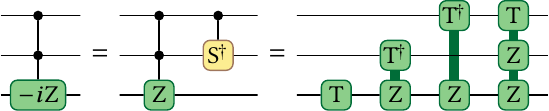}
\end{equation}
The four gates on the right hand side describe the $8\times8$ unitaries $e^{\mathrm{i}\tfrac{\pi}{8}(I \otimes I \otimes Z)}$, $e^{-\mathrm{i}\tfrac{\pi}{8}(I \otimes Z \otimes Z)}$, $e^{-\mathrm{i}\tfrac{\pi}{8}(Z \otimes I \otimes Z)}$, and $e^{\mathrm{i}\tfrac{\pi}{8}(Z \otimes Z \otimes Z)}$.  All of these four gates commute with each other.  Given a $|T\rangle$ magic state, such an operation can be implemented using multi-qubit joint ZZ measurement (\cref{fig:multi-qubit-zz-measurement}(a)) and a conditional $S^\dagger$ gate, illustrated by the implementation of $e^{\mathrm{i}\tfrac{\pi}{8}(Z\otimes Z\otimes Z)}$~\cite{BK05,LO18}:
\begin{equation}
\label{eq:exponential}
\inputtikz{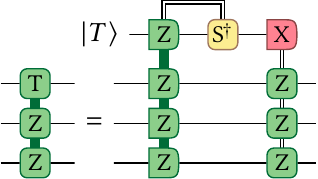}
\end{equation}
If the exponent is negative, e.g., $e^{-\mathrm{i}\tfrac{\pi}{8}(I\otimes Z\otimes Z)}$, the $S^\dagger$ gate is conditionally applied if the result of the multi-qubit joint ZZ measurement is 0.

It is possible to parellelize the execution of all exponential operators by
using the fanout gadget and creating four cat-state copies of the target qubit
and two cat-state copies for each control qubit.  More precisely we create
$z$-cat states on 5 and 3 qubits for the target qubit and each control qubit,
respectively, and then using these to both apply the exponential operators as
well as connecting the original target and control qubits via joint ZZ
measurement.  Instead of applying the $Z$ corrections in the exponential
operators on qubits $A_i$, $B_i$, and $C_i$ as suggested
by~\eqref{eq:exponential}, we collect all $X$ measurement results and
compute $Z$ corrections that are directly applied to the control and target
qubits of the operation.  For example, the $Z$ correction on the target qubit is
applied if and only if an odd number of $X$ measurement results is 1:
\begin{equation}
\label{eq:cciz-with-fanout}
\inputtikz{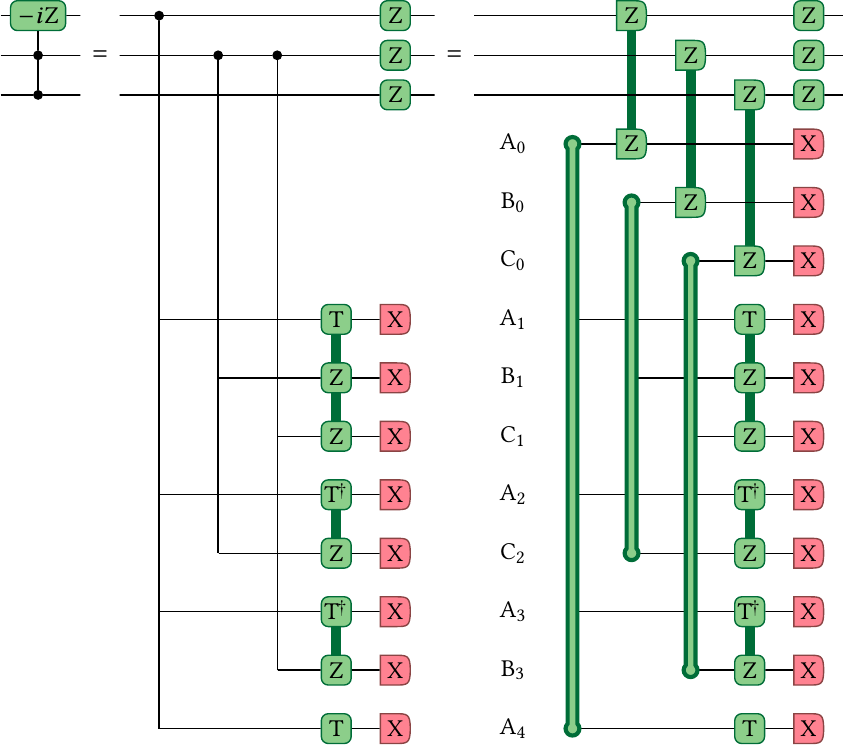}
\end{equation}
Note that we can simply transform this doubly-controlled $-\mathrm{i}Z$ circuit into a doubly-controlled $-\mathrm{i}X$ gadget by conjugating the target line with a Hadamard gate, thereby turning the corresponding joint ZZ measurement and $Z$ correction into a joint XZ measurement and $X$ correction, respectively.  The three joint measurements and Pauli corrections in the circuit correspond to those in~\cref{fig:ccix-gadget-high-level}.

The conditional $S^\dagger$ gates in the exponential operators is implemented via $|S_x\rangle$ magic states, where $|S_x\rangle = HS|+\rangle$.  We are using the circuit in~\cref{fig:multi-qubit-zz-measurement}(b) that is derived from~\cref{fig:s-injection}(a) by conjugating it with Hadamard gates and prepending a $Z$ gate since $S^\dagger = ZS$.  It is possible to use joint XZ measurement and \texttt{CNOT} gate to clone an $|S_x\rangle$ state:
\begin{equation}
\label{eq:clone-cx}
\inputtikz{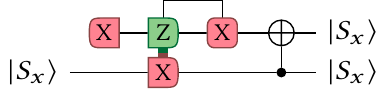}
\end{equation}
We are making use of this gadget and therefore only need to apply the $|S_x\rangle$ states before executing each \texttt{CCiX} gadget in the table lookup circuit for the first time.

We now have all building blocks to implement~\eqref{eq:cciz-with-fanout} on surface code in a $9\times 6$ grid.  As illustrated in \cref{fig:ccix-gadget-high-level}, the implementation consists of a preparation and an execution phase.  The location of key qubits in this grid is as follows:
\begin{equation}
    \label{eq:locations}
    \inputtikz{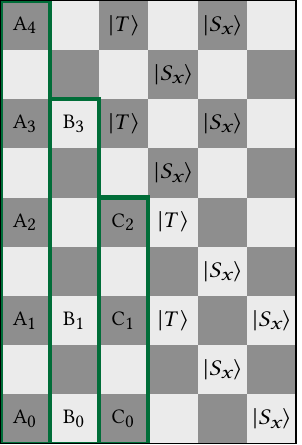}
    \qquad
    \inputtikz{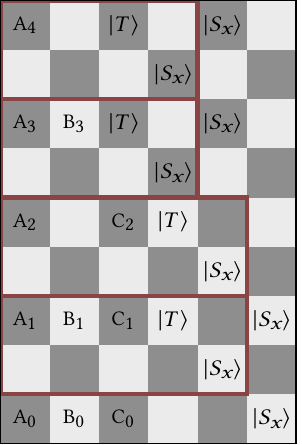}
    \qquad
    \inputtikz{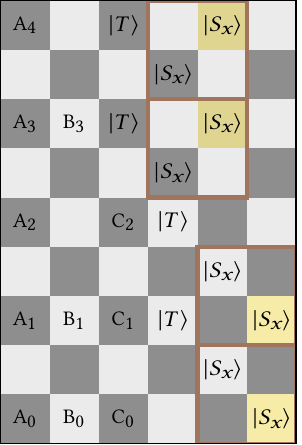}
\end{equation}
Indices $A_i$, $B_i$, and $C_i$ correspond to the cat-state copies in~\eqref{eq:cciz-with-fanout}. Cells labeled $|T\rangle$ must be prepared in a $|T\rangle$ state in the preparation phase before the exponential operation is executed in the execution phase.  Cells labeled $|S_x\rangle$ and highlighted in yellow must be prepared in an $|S_x\rangle$ state once before the \texttt{CCiX} gadget is used for the first time.  The other four cells labeled $|S_x\rangle$ will be prepared in an $|S_x\rangle$ state using the cloning gadget in~\eqref{eq:clone-cx} during each preparation stage.  Finally, areas outlined in green indicate where $z$-cat states are prepared, areas outlined in red indicate where exponential operations are executed (see~\eqref{eq:exponential}), and areas outlined in yellow indicate where $|S_x\rangle$ states are cloned (see~\eqref{eq:clone-cx}).

The preparation stage consists of the following two steps:
\begin{equation}
\label{eq:ccix-prepare}
\inputtikz{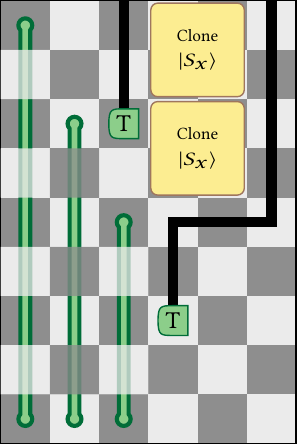}
\qquad+\qquad
\inputtikz{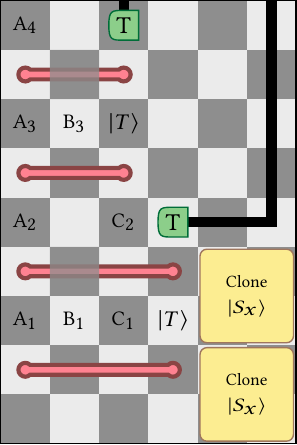}
\end{equation}
In the first step, fanout qubits are connected via $z$-cat states.  Here we use a more transparent cat state line to indicate holes in the cat state.  Also two of the four $|T\rangle$ states are prepared.  Note that a $|T\rangle$ state is delivered from the boundary (here from the top).  All qubits on the ``delivery path,'' which is indicated using the black line, must be in a blank state.  At the same time, two of the four $|S_x\rangle$ states are cloned.  In the second step, the $x$-cat states used for the multi-qubit joint $ZZ$ measurement (\cref{fig:multi-qubit-zz-measurement}) in the exponential operation (see~\eqref{eq:exponential}) are prepared and the two remaining $|T\rangle$ and $|S_x\rangle$ states are prepared and cloned, respectively.

The execution stage also consists of two steps:
\begin{equation}
\label{eq:ccix-execute}
\inputtikz{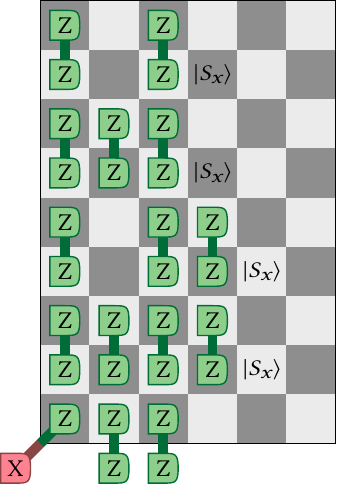}
\qquad+\qquad
\inputtikz{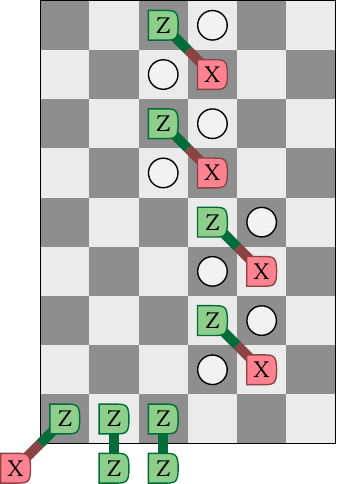}
\end{equation}
The three joint measurements in the bottom of the grid are remote measurements and combine the interface qubits $A_0$, $B_0$, and $C_0$ (see~\eqref{eq:cciz-with-fanout}) with the original target qubit and control qubits, respectively.  All other joint measurements are on adjacent qubits.  The remote measurements in both steps are the same and their execution time is expected to dominate that of local joint ZZ and XZ measurements.

\bibliographystyle{ACM-Reference-Format}
\bibliography{library,references}

\newpage
\appendix
\section{Extended set of surface code operations}

\label{sec:extended-operations}

\subsection{Axis-independent operations}
\label{sec:axis-independent-operations}

\begin{figure}
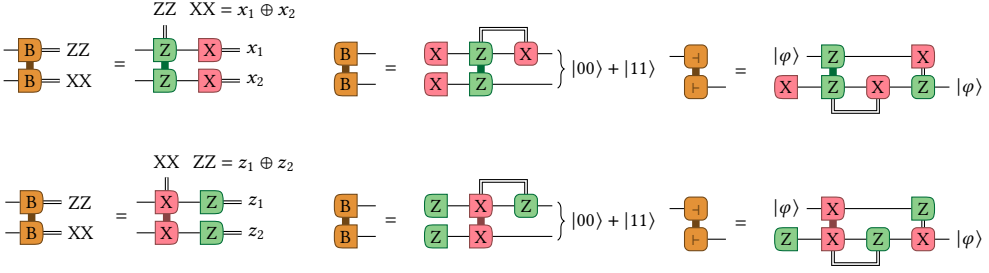

    \centering
    \inputtikz{axis-independent-operations}
    \caption[Axis independent operations]{
    Axis independent Bell measurement, Bell state preparation and Move operations.
    }
    \label{fig:axis-independent-operations}
\end{figure}

It is tedious to always have to deal with horizontal and vertical directions for XX and ZZ measurements, 
so the first step is to introduce axis-independent gadgets of commonly-used operations. In~\cref{fig:axis-independent-operations}, we present axis-independent circuits for Bell measurement, Bell preparation, and the move operation, which can be used to move the state of a qubit to an ancillary qubit.

\subsection{Remote Bell measure and Bell preparation}
\label{sec:remote-bell-measure-and-preparation}

A remote operation is a constant depth circuit implementing 
a unitary or a measurement operation between several target qubits connected by a path of 
ancillary qubits. We make heavy use of such operations in our implementation of table lookup.

\begin{figure}
    \centering
    \subfloat[Quantum teleportation]{\inputtikz{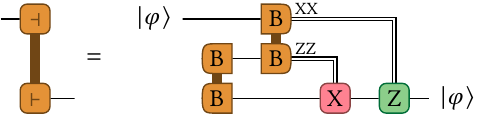}}
    \hfill
    \subfloat[Remote Bell preparation]{\inputtikz{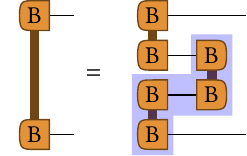}\qquad\qquad}
    \hfill
    \subfloat[Remote Bell measurement]{\inputtikz{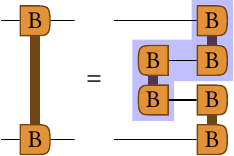}\qquad\qquad\qquad}

    \caption{(a) shows a teleportation circuit composed of Bell state preparation and Bell measurement. (b) and (c) show how to extend the teleportation circuit, highlighted in blue, to construct remote Bell state preparation and Bell measurement.}
    \label{fig:teleport-bell-measure-and-prep}
\end{figure}

\cref{fig:teleport-bell-measure-and-prep} shows two examples of simple remote operations
derived from a teleportation circuit. The diagram shows Bell preparation and Bell measurement on two qubits that are
connected by a path consisting of two ancillary qubits. Replacing local Bell measurement or Bell preparation with 
the remote version allows us to perform Bell measurements between two arbitrary qubits that are connected by a path of even length in constant depth. To support constant-depth remote Bell measurements with paths of odd lengths, we can make use of the move operation from \cref{fig:axis-independent-operations}.

\subsection{Remote ZZ and XX measurements}
\label{sec:remote-zz-and-xx-measurements}

\begin{figure}[t]
    \centering
    \footnotesize
    \scalebox{.95}{\inputtikz{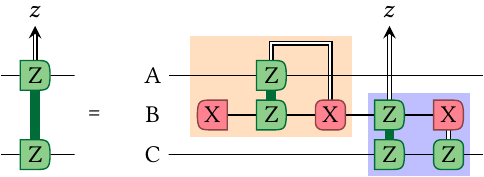}}
    \hfill
    \scalebox{.95}{\inputtikz{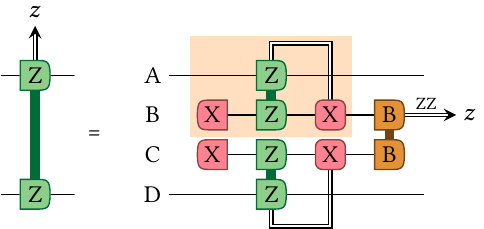}}
    \hfill
    \scalebox{.95}{\inputtikz{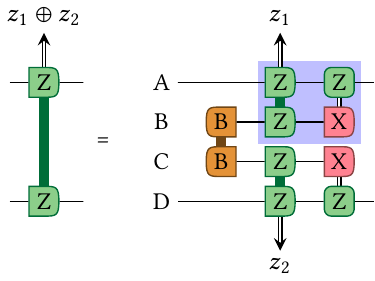}}
    \caption[Remote ZZ measurements on three or four qubits]{
    Remote ZZ measurements on three or four qubits.
    The orange rectangle highlights ``growth along Z direction,'' whereas the 
    blue rectangle indicates a ``contraction along Z direction.''
    It is possible that these gadgets can become an optimized operation available directly from surface code.
    }
    \label{fig:remote-zz-measurement-three-four-qubits}
\end{figure}

Equipped with gadgets for remote Bell measurement and Bell state preparation, we can construct circuits for remote ZZ and XX measurements.
As examples for remote joint ZZ measurement, we show three- and four-qubit measurement circuits in~\cref{fig:remote-zz-measurement-three-four-qubits}.

The three-qubit circuit works as follows: The gates in the orange box fan out qubit A to qubit B. The joint ZZ measurement in the blue box is then equivalent to a joint ZZ measurement on qubits A and C. After this joint ZZ measurement, a conditional X-correction could be applied to qubit B in order to ensure that qubits B and C are in a +1-eigenstate of ZZ, allowing us to uncompute qubit B through an X-measurement (and a conditional Z fixup on qubit C). Note, however, that the X-correction on qubit~B can be omitted since it commutes with the subsequent X-measurement.

The first four-qubit circuit in \cref{fig:remote-zz-measurement-three-four-qubits} fans out qubits A and D to qubits B and C, respectively, before performing a Bell measurement across B and C, which yields the result of the joint ZZ measurement. In contrast, the second four-qubit circuit first establishes a Bell state on qubits B and C. The top joint ZZ measurement then connects the Bell state to qubit A, resulting in qubit A fanned-out across B and C (with a conditional Pauli-X gate on qubits B and C), whereas the bottom joint ZZ measurement can be seen as actually jointly measuring qubits A and D (keeping in mind the potential X-corrections on B and C from the first joint ZZ measurement). The result of the joint ZZ measurement of qubits A and D is thus given by the XOR of the two joint ZZ measurements in the circuit. As was the case for the three-qubit circuit, since---up to conditional X-corrections---qubits A and D are fanned out to qubits B and C, respectively, B and C can be uncomputed using an X-measurement (which commutes with all conditional corrections) and a conditional Z-correction on qubits A and D.

\begin{figure}
    \caption[Remote ZZ measurements timings]{
    Remote ZZ measurement on five qubits and timing of multi-qubit remote joint measurement operations.
    The orange rectangle highlights the ``grow along Z direction'' gadget, whereas the
    blue rectangle highlights the ``contraction along Z direction'' gadget.
    It is possible that these gadgets can become an optimized operation available directly from surface code. Time $\tau_1$ is the maximum of the times required for Bell state preparation and ``growing along Z direction.''
    Time $\tau_2$ is the maximum of the times required for a Bell measurement and ``contraction along Z direction.''
    }
    \label{fig:remote-zz-measurements}
    \centering
    \footnotesize
    \inputtikz{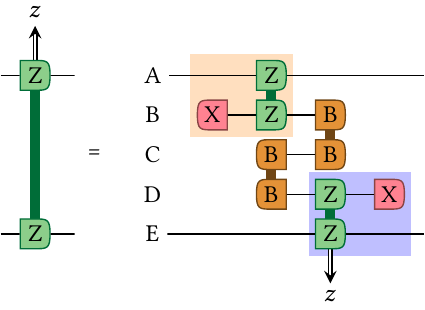}    
    \hfill
    \inputtikz{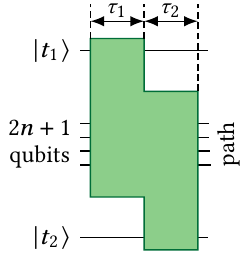}
    \quad
    \inputtikz{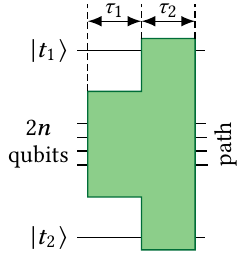}
    \quad
    \inputtikz{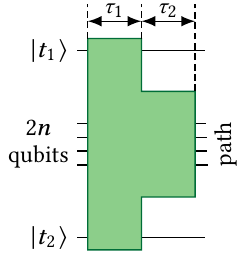}
\end{figure}

These examples can be extended to 
a general remote ZZ measurement along an arbitrary path. 
For example, we get the five-qubit circuit in \cref{fig:remote-zz-measurements} by inserting the teleportation circuit from~\cref{fig:teleport-bell-measure-and-prep} into the three-qubit circuit between steps three and four, thus teleporting qubit B to qubit D.
Every circuit on $k$ qubits for remote ZZ measurement can be made into circuit on $k+2$ qubit 
by replacing two-qubit Bell preparation or measurement circuits with the four-qubit remote Bell preparation or Bell measurement circuits in~\cref{fig:teleport-bell-measure-and-prep}.
In~\cref{fig:remote-zz-measurement-three-four-qubits}, we present two circuits on an even number of qubits that we derived from two different four-qubit circuits. Similarly, we derive the circuits for an odd number of qubits from the five-qubit circuit in~\cref{fig:remote-zz-measurements}.
Possible timing shapes for remote ZZ measurements are shown in \cref{fig:remote-zz-measurements}.
Remote XX measurement circuits can be derived from remote ZZ measurement circuits via a simple rule: all Xs need to be replaced by Zs and all Zs need to be replaced by Xs in the circuit. 
Bell measurements and Bell preparations are left unchanged.
Finally, we note that remote ZZ measurement can be performed along an arbitrary path, as long as the target qubits are vertically-adjacent to their ancillary neighbors. Analogously, for remote XX we require the target to be horizontally-adjacent to its ancillary neighbors.

\paragraph*{Remote joint measurement with move.}
Remote joint XX and remote joint ZZ measurements can be constructed on an almost
arbitrary path of consecutive adjacent qubits.  However, the first pair of
qubits and the last pair of qubits must be either horizontally adjacent, in case
of a remote XX measurement, or vertically adjacent, in case of a remote ZZ
measurement.  We will see that gadgets described in the remainder of this
section often immediately perform a single qubit measurement on one of the
endpoints of a remote joint measurement.  In that case we can leverage
teleportation to relax the adjacency requirement on one of the endpoints of the
remote joint measurement, e.g., illustrated for a remote ZZ measurement:
\begin{equation}
\label{eq:remote-zz-with-move}
\inputtikz{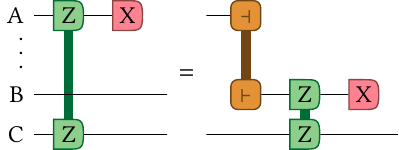}
\end{equation}
Here, qubits A and B are not adjacent, but qubits B and C are.  Note that the
joint ZZ measurement can be executed at the same time as the Bell measurements
in the teleportation circuit.

\subsection{Remote CNOT}
\label{sec:remote-cx}

\begin{figure}[t]
    \centering
    \scalebox{.95}{\inputtikz{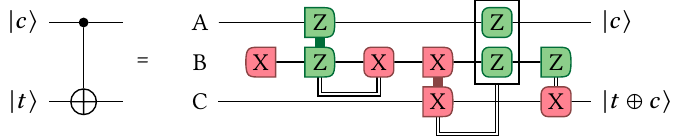}}
    \scalebox{.95}{\inputtikz{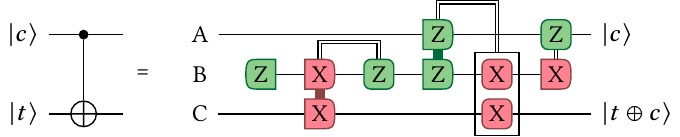}}
    
    \bigskip
    \scalebox{.95}{\inputtikz{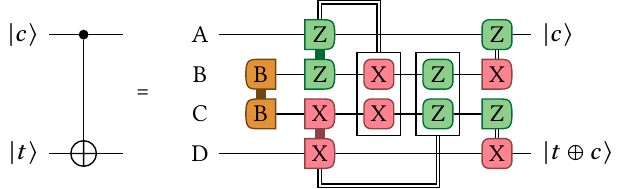}}
    \qquad
    \scalebox{.95}{\inputtikz{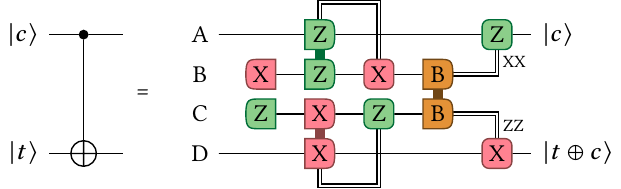}}

    \caption[CNOT on three and four qubits]{
    CNOT on three or four qubits.
    It is possible that these gadgets can become an optimized operation available directly from surface code.
    }
    \label{fig:cx-three-four-qubits}
\end{figure}

\begin{figure}
    \caption[Remote CNOT timings]{
    Remote CNOT gate on five qubits and timing of multi-qubit remote CNOT operations.
    It is possible that these gadgets can become an optimized operation available directly from surface code.  The orange rectangles highlight the ``grow along X or Z direction'' gadget, whereas the blue rectangles highlight the ``contraction along X or Z direction'' gadgets.  Time $\tau_1$ is the maximum of the times required for Bell state preparation and ``growing along X or Z direction.''  Time $\tau_2$ is the maximum of the times required for Bell measurement and ``contraction along X or Z direction.''}
    \label{fig:remote-cx}
    \centering
    \inputtikz{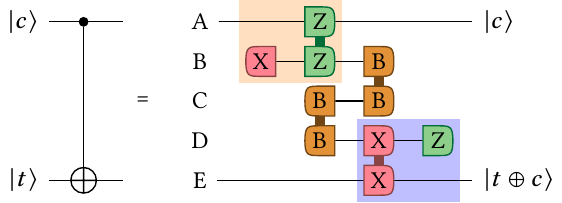}
    \qquad
    \inputtikz{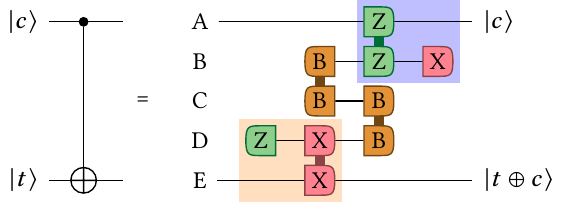}
    
    \bigskip
    \inputtikz{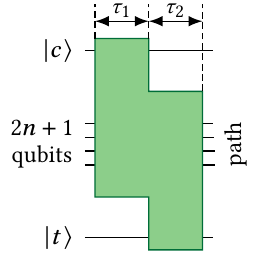}
    \quad
    \inputtikz{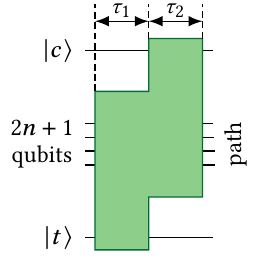}
    \quad
    \inputtikz{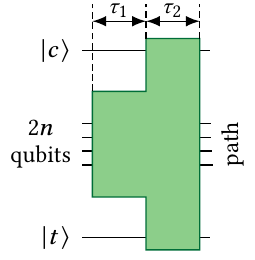}
    \quad
    \inputtikz{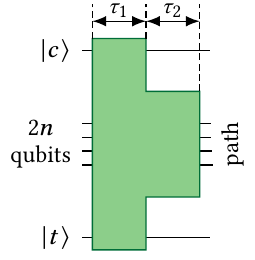}
\end{figure}

Here, we present several constructions for executing a CNOT gate.
We start by presenting examples of a CNOT on three and four qubits in \cref{fig:cx-three-four-qubits} (see also~\cite{FG19}).
The four-qubit circuit can be derived from the three-qubit circuit by replacing the local ZZ measurement 
with the three-qubit ZZ measurement circuit from~\cref{fig:remote-zz-measurement-three-four-qubits} and by
using the Bell preparation and measurement identities in~\cref{fig:teleport-bell-measure-and-prep}.
We get the five-qubit circuits for CNOT in \cref{fig:remote-cx} by inserting the teleportation circuit from~\cref{fig:teleport-bell-measure-and-prep} into the three-qubit circuits for CNOT between steps three and four, thus teleporting qubit D to qubit B.
Every circuit on $k$ qubits for remote CNOT can be made into circuit on $k+2$ qubit 
by replacing two-qubit Bell preparation or measurement with four qubit remote Bell preparation or Bell measurement circuits in~\cref{fig:teleport-bell-measure-and-prep}.
We derive two circuits on even number of qubits from two different four qubit circuits in~\cref{fig:cx-three-four-qubits}. Similarly we derive odd number of qubit circuits from the five qubit circuits in~\cref{fig:remote-cx}.
Possible timing shapes for remote CNOTs are shown in \cref{fig:remote-cx}.  Finally, we note that remote CNOTs can be performed along arbitrary path, as long as the target qubit is horizontally adjacent to its ancillary neighbour and the control qubit is vertically adjacent to its ancillary neighbour.

\subsection{Remote XZ measurement}
\label{sec:remote-zx-measurement}
\begin{figure}[t]
    \centering
    \footnotesize
    \centering
    \scalebox{.9}{\inputtikz{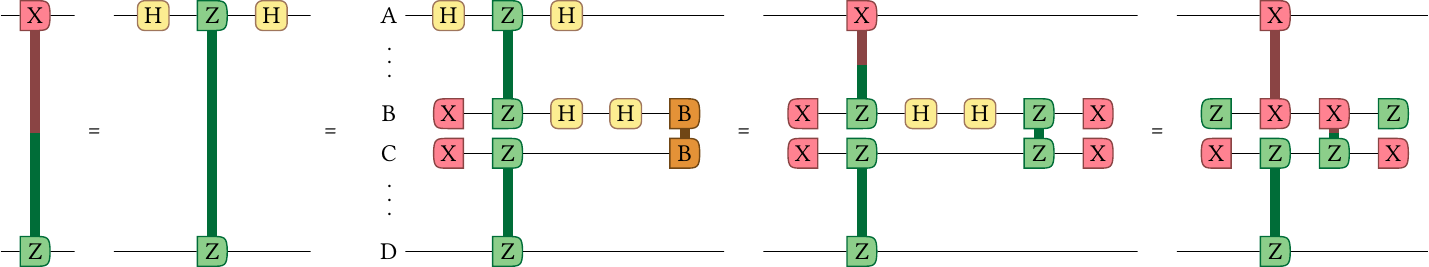}}
    \caption{This example shows how to derive XZ measurement from two remote ZZ measurements combined by Bell measurement.  By commuting Hadamard gates we change the topmost remote ZZ measurement and the Bell measurement into a remote XX measurement and an XZ measurement, respectively.}
    \label{fig:my_label}
\end{figure}
We can construct a remote XZ measurement by replacing the topmost or lowermost joint ZZ measurement with a joint XZ measurement.  We can also arbitrarily changing an inner joint ZZ measurement by propagating Hadamard gates through a circuit for remote ZZ measurement.  Figure~18 illustrates this based on a circuit that combines two remote ZZ measurements with Bell measurement (see also Fig.~\ref{fig:remote-zz-measurement-three-four-qubits}).  The Hadamard gates can be merged into Prepare X and Measure X operations, thereby turning the first remote ZZ measurement into a remote XX measurement and the joint ZZ measurement from the Bell measurement circuit into an XZ measurement. 

\subsection{Cat state preparation}
\label{sec:cat-state-preparation}
We use recursive constructions to prepare the states
\begin{equation}
    \label{eq:ghz}
    |\mathrm{GHZ}_n\rangle_Z = |\mathrm{GHZ}_n\rangle = \tfrac{1}{\sqrt2}(|0\rangle^{\otimes n} + |1\rangle^{\otimes n})
\end{equation}
and
\begin{equation}
    \label{eq:ghx}
    |\mathrm{GHZ}_n\rangle_X = H^{\otimes n}|\mathrm{GHZ}_n\rangle = \tfrac{1}{\sqrt2}(|+\rangle^{\otimes n} + |-\rangle^{\otimes n})
\end{equation}
on $n$ vertically adjacent and $n$ horizontally adjacent qubits, respectively.  We also refer to these states as $z$-cat states and $x$-cat states.  Note that
\[
|\mathrm{GHZ}_n\rangle_X = \frac{1}{\sqrt{2^{n-1}}}\sum\limits_{\substack{0\le i < 2^n \\ \nu i \bmod 2 = 0}}|i\rangle,
\]
i.e., the uniform superposition over all basis states with an even number of 1s in their binary representation.\footnote{We use $\nu i$ to refer to the sideways sum of $i$, i.e., the number of 1s in $i$'s binary representation.}

We describe our construction for $|\mathrm{GHZ}_n\rangle_Z$ first; the construction for $|\mathrm{GHZ}_n\rangle_X$ is similar.  The construction is based on the following decomposition of cat states.  For any $n \ge 2$ and $j + k = n$ with $1 \le j < n$, we have
\[
  \begin{aligned}
  |\mathrm{GHZ}_j\rangle \otimes |\mathrm{GHZ}_k\rangle
  & = \tfrac{1}{2}(
  |0^j0^k\rangle + |0^j1^k\rangle + |1^j0^k\rangle + |1^j1^k\rangle
  )
  \end{aligned}
\]
We can now measure qubit $j$ (the last qubit of the first cat state) and qubit $j+1$ (the first qubit of the second cat state) using a joint ZZ measurement.  If the measurement outcome corresponds to the +1 eigenvalue, the resulting state is $|\mathrm{GHZ}_n\rangle$ and otherwise, the resulting state is $\frac{1}{\sqrt2}(|0^j1^k\rangle + |1^j0^k\rangle)$, which can be transformed into $|\mathrm{GHZ}_n\rangle$ by applying $X$ corrections on either the first $j$ or the last $k$ qubits.

\begin{figure}[t]
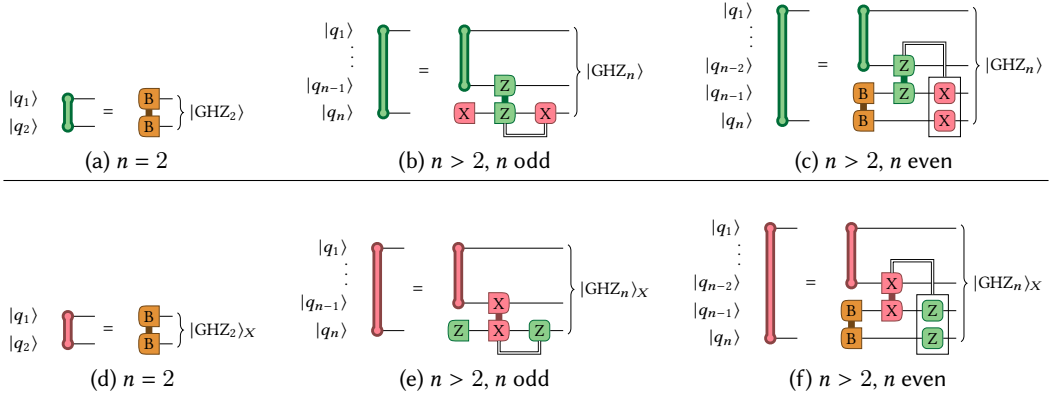

    \caption{Cat state preparation on 2 or more vertically adjacent qubits.  The cat state preparations used on the right-hand sides are applied to an even number of qubits.  Further, consecutive Pauli corrections can be merged by computing the parity of respective measurement results.}
    \label{fig:cat-state}
    \footnotesize
    \medskip
    
    \subfloat[$n=2$]{\scalebox{.9}{\inputtikz{ghz-2}}}
    \hfill
    \subfloat[$n>2$, $n$ odd]{\scalebox{.9}{\inputtikz{ghz-n-odd}}}
    \hfill
    \subfloat[$n>2$, $n$ even]{\scalebox{.9}{\inputtikz{ghz-n-even}}}
    
    \hrulefill
    \medskip
    
    \subfloat[$n=2$]{\scalebox{.9}{\inputtikz{ghx-2}}}
    \hfill
    \subfloat[$n>2$, $n$ odd]{\scalebox{.9}{\inputtikz{ghx-n-odd}}}
    \hfill
    \subfloat[$n>2$, $n$ even]{\scalebox{.9}{\inputtikz{ghx-n-even}}}
\end{figure}

Figure~\ref{fig:cat-state}(a)--(c) shows our construction based on this decomposition, in which $k \in \{1, 2\}$ such that $j$ is even and $X$ corrections are applied on the last $k$ qubits.  The base case is  $|\mathrm{GHZ}_2\rangle = \frac{1}{\sqrt2}(|00\rangle + |11\rangle)$ when $j = 2$.  Since the cat state preparations used in the recursion are applied on an even number of qubits, the resulting circuits have constant depth, with the first layer consisting of Bell state preparations and the second layer consisting of joint ZZ measurements.

Figure~\ref{fig:cat-state}(d)--(f) shows how to create $|\mathrm{GHZ}_n\rangle_X$ states analogously by replacing $X$ preparation, joint ZZ measurements, and $X$ corrections, by $Z$ preparation, joint XX measurements, and $Z$ corrections, respectively.

\paragraph*{Cat states with holes.}  Let $Q = \{q_1, \dots, q_n\}$ be a set of vertically adjacent qubits and let $H = \{h_1, \dots, h_m\} \subseteq Q \setminus \{q_1\}$.  Then we can prepare a $z$-cat state on qubits $Q \setminus H$ by first preparing a $z-$cat state on $Q$ using the construction above, followed by $X$ measurements on all qubits in $H$, resulting in measurements $r_1, \dots, r_m$.  We conditionally apply a $Z$ correction on $q_1$, if $r_1 \oplus \cdots \oplus r_m$ is true.  The technique can also be used to prepare an $x$-cat state on horizontally adjacent qubits, by replacing $X$ measurements and $Z$ corrections with $Z$ measurements and $X$ corrections, respectively.

\paragraph*{Cat states on arbitrary paths.}
We now show how to create a $z$-cat state on an arbitrary path by leaving holes on segments of horizontally adjacent qubits.  The technique works analogously for $x$-cat states.  Contrary to the previous construction, we cannot first create the cat state since joint ZZ measurements can only be applied to vertically adjacent qubits.

Let $Q = (q_1, \dots, q_n)$ be a tuple of qubits, where $q_i$ and $q_{i+1}$ are either vertically or horizontally adjacent.  As notation, let $\mathrm{vadj}(q, q')$ and $\mathrm{hadj}(q, q')$ be predicates for the respective adjacency property.  We partition $Q$ into
\begin{equation}
\label{eq:cat-partition}
q_1 \mathbin{|} q_2, q_3 \mathbin{|} q_4, q_5 \mathbin{|} \dots,
\end{equation}
where $q_n$ is part of a pair only if $n$ is odd.  Then, qubits in $Q$ are part of the cat state if they are not part of a pair, or if they are part of a pair that is vertically adjacent.  This is formally captured by the predicate
\[
\mathrm{incat}(q_i) = [i = 1] \lor ([n \bmod 2 = 0] \land [i = n]) \lor \mathrm{vadj}(q_{2\lfloor\frac{i}{2}\rfloor}, q_{2\lfloor\frac{i}{2}\rfloor+1}).
\]
Let $C = \{q \in Q \mid \mathrm{incat}(q)\}$.  Then we can construct a cat state on $C$ in $Q$ with holes $H = Q \setminus C$ using the following procedure:
\begin{enumerate}
    \item Prepare Bell pairs on qubits $q_{2i-1}$ and $q_{2i}$ for $1 \le i < \frac{n}{2}$. If $n$ is even, prepare qubit $q_n$ in a $|+\rangle$ state.
    \item For each pair $q_i, q_{i+1}$ as in the partition of~\eqref{eq:cat-partition}, do the following based on its adjacency.
    
    If $\mathrm{hadj}(q_i, q_{i+1})$, measure the pair of qubit in Bell basis and retrieve two measurement results $\mathrm{ZZ}$ and $\mathrm{XX}$ (see Fig.~\ref{fig:axis-independent-operations}).  Apply a $Z$ correction to $q_1$ if $\mathrm{XX}$ is true, and apply an $X$ correction to $q_{i+2}$ if $\mathrm{ZZ}$ is true.
    
    If $\mathrm{vadj}(q_i, q_{i+1})$, perform a joint ZZ measurement on $q_i$ and $q_{i+1}$.  If the measurement result is true, apply $X$ corrections to all qubits in $C \cap \{q_1, \dots, q_i\}$.
\end{enumerate}

\begin{figure}[t]
\caption{This cat state along an arbitrary path enables creating a longer $z$-cat state in a rectangular shape using a snake line pattern.}
\label{fig:cat-arb-path}
\centering
\scalebox{.85}{\inputtikz{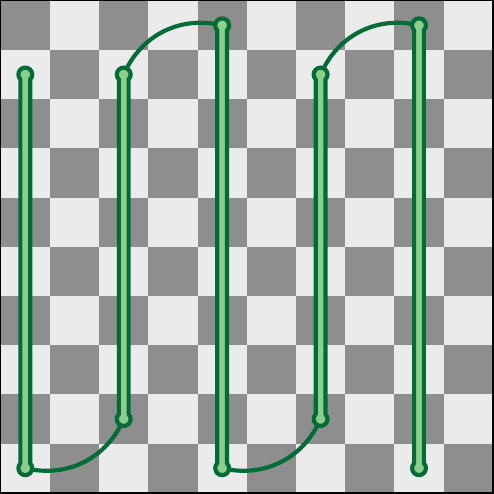}}$=$
\scalebox{.85}{\inputtikz{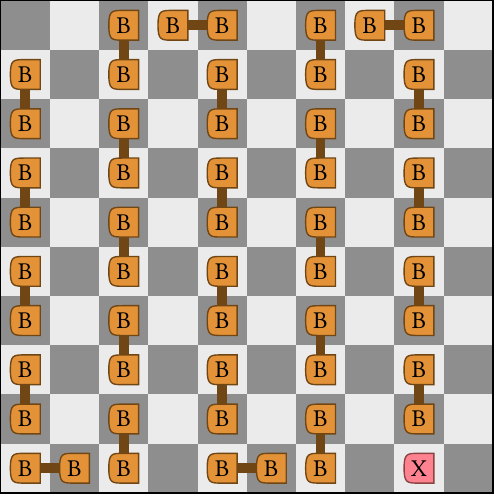}}$+$
\scalebox{.85}{\inputtikz{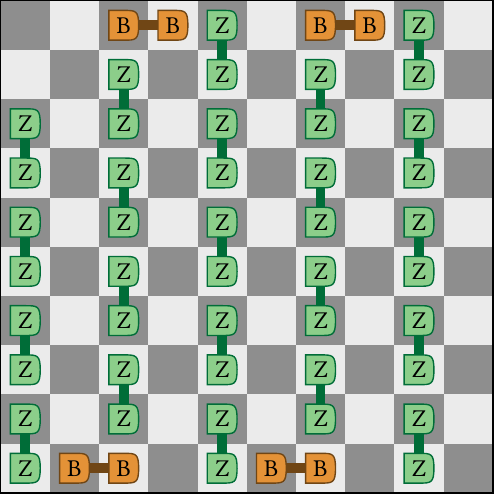}}
\end{figure}

Figure~\ref{fig:cat-arb-path} illustrates an example using this construction in a rectangular shape.

\paragraph*{Quantum fanout.}  The quantun fanout operation achieves the mapping
\begin{equation}
    \label{eq:fanout}
    (\alpha\ket{0}+\beta\ket{1}) \otimes |0^{n-1}\rangle \mapsto
    \alpha|0^n\rangle + \beta|1^n\rangle.
\end{equation}
This operation can be implemented in constant depth by first creating a $z$-cat state on the latter $n-1$ qubits, which results in
\[
  \tfrac{\alpha}{\sqrt2}|0^n\rangle + \tfrac{\beta}{\sqrt2}|1^n\rangle
  + \tfrac{\alpha}{\sqrt2}|01^{n-1}\rangle + \tfrac{\beta}{\sqrt2}|10^{n-1}\rangle.
\]
Similarly as done in the decomposition above, by applying a joint ZZ measurement on the first two qubits, one either obtains the expected result when the measurement outcome corresponds to the +1 eigenvalue or, otherwise, an $X$-correction on each of the last $n-1$ qubits is needed.  Since the additional joint ZZ measurement nicely fits into the layer of joint ZZ measurements for preparing the cat state, the execution time of applying quantum fanout equals the execution time of preparing a cat state.

\subsection{Multi-target CNOT}
\label{sec:multi-cnot}

Table lookup relies on a multi-target CNOT to write the output bits into the target qubits.
We thus present a resource-efficient implementation (based on low-depth fanout) in terms of our chosen low-level gate set.

\begin{figure}[t]
    \centering
    \scalebox{.75}{\footnotesize\inputtikz{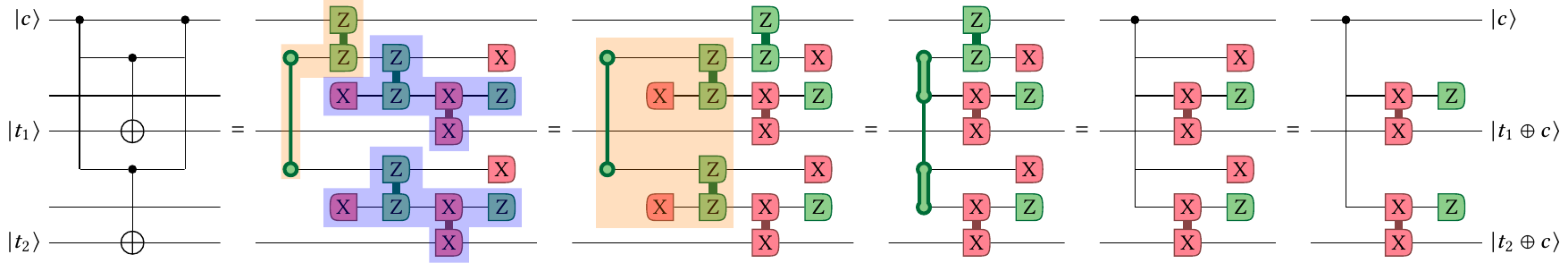}}
    \caption{Derivation of multi-target CNOT based on quantum fanout and CNOT gates.}
    \label{fig:mtx-derivation}
\end{figure}

We depict a step-by-step derivation of our multi-target CNOT gate in Figure~\ref{fig:mtx-derivation}. Starting with an abstract quantum circuit using fanout (see Section~\ref{sec:cat-state-preparation}) and CNOT gates (see Figure~\ref{fig:cx-three-four-qubits}) on the left-hand side of the figure, we first decompose these operations into surface code primitives. The low-level primitives corresponding to quantum fanout and CNOT gates are highlighted in orange and blue, respectively.  In the second step, we move the first joint ZZ measurement past the joint ZZ measurement from the first CNOT gate.  We then note that we can extend the two-qubit $z$-cat state into a four qubit $z$-cat state (highlighted in orange in the third circuit diagram).  This four-qubit cat state together with the joint ZZ measurement on the top two qubits is a quantum fanout operation.  Two of the four fanout targets are immediately removed by the $X$ measurement, and therefore the corresponding qubits can be removed altogether, resulting in the final circuit diagram of our multi-target CNOT gate. 

In a 2D grid setting, we can use qubits that are horizontally adjacent to cat state qubits as target qubits.  Here is an example using a simple vertical cat state:
\[
\inputtikz{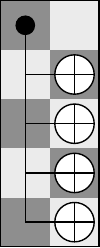}=
\inputtikz{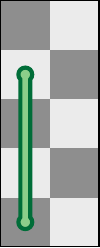}+
\inputtikz{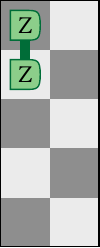}+
\inputtikz{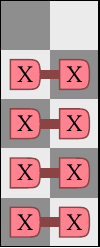}+
\inputtikz{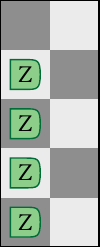}
\]
Note that the joint ZZ measurement and the cat state preparation can be performed simultaneously, resulting in the control qubit being fanned out to all qubits in the left column before the joint XX measurements and the final Z-measurements are applied.

\end{document}